\documentclass[journal]{IEEEtran}
\usepackage{xcolor}
\usepackage{graphicx}

\usepackage{bm}

\usepackage{amsmath}
\usepackage{subfigure}
\usepackage{amsthm}
\usepackage{amsmath, amssymb}
\usepackage{cite}

\newcommand{\rev}[1]{\textcolor{black}{#1}}
\newcommand{\revref}[1]{\textcolor{black}{#1}}

\begin{document}
\title{A Survey: Collaborative Hardware and Software \\ Design in the Era of Large Language Models}
%
%
%

\author{Cong~Guo,
Feng~Cheng,
Zhixu~Du,
James~Kiessling,
Jonathan~Ku,
Shiyu~Li,
Ziru~Li,
Mingyuan~Ma,
Tergel~Molom-Ochir,
Benjamin~Morris,
Haoxuan~Shan,
Jingwei~Sun,
Yitu~Wang,
Chiyue~Wei,
Xueying~Wu,
Yuhao~Wu,
Hao~Frank~Yang,
Jingyang~Zhang,
Junyao~Zhang,
Qilin~Zheng,
Guanglei~Zhou,
Hai~(Helen)~Li,~\IEEEmembership{Fellow,~IEEE,}
and~Yiran~Chen,~\IEEEmembership{Fellow,~IEEE}
\thanks{This work was supported in part by the NSF under Grant 2328805 and Grant 2112562, and ARO W911NF-23-2-0224. (Corresponding author: Yiran Chen.)}
\thanks{Cong~Guo,
Feng~Cheng,
Zhixu~Du,
James~Kiessling,
Jonathan~Ku,
Shiyu~Li,
Ziru~Li,
Mingyuan~Ma,
Tergel~Molom-Ochir,
Benjamin~Morris,
Haoxuan~Shan,
Jingwei~Sun,
Yitu~Wang,
Chiyue~Wei,
Xueying~Wu,
Yuhao~Wu,
Jingyang~Zhang,
Junyao~Zhang,
Qilin~Zheng,
Guanglei~Zhou,
Hai~(Helen)~Li,
and~Yiran~Chen are with the Department of Electrical and Computer Engineering, Duke University, Durham,
NC, 27705, USA.}
\thanks{
Hao~Frank~Yang is with the Department of Civil and Systems Engineering, Johns Hopkins University, Baltimore, Maryland, 21218, USA.}
}

%
%

\markboth{Journal of \LaTeX\ Class Files,~Vol.~14, No.~8, August~2015}%
{Shell \MakeLowercase{\textit{et al.}}: Bare Demo of IEEEtran.cls for IEEE Journals}
%


\IEEEspecialpapernotice{(Invited Paper)}

\maketitle
\begin{abstract}
The rapid development of large language models (LLMs) has significantly transformed the field of artificial intelligence, demonstrating remarkable capabilities in natural language processing and moving towards multi-modal functionality. 
These models are increasingly integrated into diverse applications, impacting both research and industry. 
However, their development and deployment present substantial challenges, including the need for extensive computational resources, high energy consumption, and complex software optimizations. 
Unlike traditional deep learning systems, LLMs require unique optimization strategies for training and inference, focusing on system-level efficiency. 
This paper surveys hardware and software co-design approaches specifically tailored to address the unique characteristics and constraints of large language models.
This survey analyzes the challenges and impacts of LLMs on hardware and algorithm research, exploring algorithm optimization, hardware design, and system-level innovations. 
It aims to provide a comprehensive understanding of the trade-offs and considerations in LLM-centric computing systems, guiding future advancements in AI. 
Finally, we summarize the existing efforts in this space and outline future directions toward realizing production-grade co-design methodologies for the next generation of large language models and AI systems.
\end{abstract}

\section{Introduction}
The rapid advancement of large language models~\cite{gpt,touvron2023llama,team2023gemini} (LLMs) has brought revolutionary change to the landscape of artificial intelligence (AI). 
These sophisticated models, leveraging vast amounts of data and significant computational power, have pushed the boundaries of what AI systems can achieve, demonstrating unprecedented capabilities in natural language understanding, generation, and interaction. 
Furthermore, LLMs are progressing by incorporating tasks beyond natural language processing, moving towards achieving multi-modal functionality. 
As LLMs become increasingly integrated into a wide range of applications—from chatbots~\cite{gpt-3,gpt-4,touvron2023llama,llama2} and virtual assistants~\cite{liu2023llava,gpt-4} to complex decision-making systems—their impact on research and industry becomes increasingly profound.

\rev{
Despite their success in various application fields, 
LLMs face unique challenges compared to CNN models, particularly in \textbf{training} and \textbf{inference}.
Due to their vast number of parameters, often in the billions or even trillions, LLMs require significantly more memory during training. 
For example, training a model like GPT-3~\cite{gpt-3}, which has 175 billion parameters, demands around 350GB of GPU memory just for storing model parameters. 
In contrast, a typical CNN such as ResNet-50~\cite{resnet}, with 25 million parameters, requires only about 100MB of memory for weights. 
This vast difference in memory requirements makes training LLMs much more demanding. 
Solutions to address this include model parallelism, which splits the model across multiple devices to distribute memory usage; mixed-precision training, which reduces memory consumption by using lower-precision data types; and memory-efficient optimizers like DeepSpeed's ZeRO~\cite{zero}, which reduces the memory footprint during training.
}

\rev{
In terms of inference, LLMs are inherently larger and require more computational power and memory than CNNs. 
This makes deploying LLMs significantly more resource-intensive. The autoregressive nature of LLMs also exacerbates the memory wall~\cite{gholami2024ai} problem because each token generated depends on all previously generated tokens, resulting in increased memory and computational requirements as the sequence length grows. 
This differs from convolutional neural networks (CNNs), where computations can be parallelized more efficiently. 
Furthermore, LLMs use key-value (KV) caches to store activations from previous tokens, speeding up subsequent token generation but also necessitating the storage of large amounts of activation data. 
As the sequence length increases, the KV cache grows linearly, posing significant memory management challenges, especially for longer contexts.
} 

In addition to challenges, LLMs also offer unique opportunities for improved efficiencies. 
Unlike CNNs, which employ diverse operators, LLMs have similar architectures. This consistency allows for the custom-made implementation of operators specific to certain architectures or hyperparameters.

This survey aims to analyze the unique challenges posed by LLMs and their significant impact on research directions within both the hardware and algorithm communities. We examine existing works on algorithm optimization, hardware architecture design, and system-level innovations for LLMs. Through this survey, we strive to develop a comprehensive understanding of the intricate trade-offs and design considerations that govern the development of LLM-centric computing systems. By synthesizing the latest research findings and identifying emerging trends, we aim to pave the way for future breakthroughs in this rapidly evolving field, enabling the creation of more powerful and efficient artificial intelligence systems.

The structure of the remaining survey is as follows: 
Section~\ref{sec:bg} introduces the preliminary knowledge related to LLMs.
In Section~\ref{sec:train}, we examine the current best practices for LLM training. 
In Section~\ref{sec:inferece}, we discuss the latest hardware and software co-design techniques for LLM inference.
Finally, Section~\ref{sec:conclusion} summarizes the main contributions of this survey.

\section{Preliminaries}
\label{sec:bg}

Large Language Models (LLMs) leverage massive datasets and sophisticated architectures to understand, generate, and manipulate human language with unprecedented accuracy and fluency. The backbone of modern LLMs is the transformer architecture, which has revolutionized NLP by addressing the limitations of previous recurrent and convolutional models.

\subsection{Transformer Architecture}
The transformer architecture, introduced by Vaswani et al. in the paper ``Attention is All You Need,"~\cite{vaswani2017attention} consists of an encoder-decoder structure. 
However, many LLMs, like GPT~\cite{gpt,gpt-2,gpt-3,gpt-4} (Generative Pre-trained Transformer), use only the decoder part. 
The core innovation of the transformer is the multi-head self-attention mechanism~\cite{vaswani2017attention} (MHSA), which enables the model to weigh the importance of different words in a sentence. 

\subsubsection*{Linear Projection}
In the MHSA block, input embeddings are first linearly projected into three different spaces to generate queries (Q), keys (K), and values (V). 
These projections are performed through learned linear transformations, which means that the input embeddings are multiplied by different weight matrices to produce Q, K, and V. 
Mathematically, this can be expressed as:
$$
Q = XW_Q, \quad K = XW_K, \quad V = XW_V
$$
where \( X \) represents the input embeddings, and \( W_Q \), \( W_K \), \( W_V \) are the learned weight matrices for the queries, keys, and values, respectively. 
Each head in the multi-head attention mechanism independently performs this projection, enabling the model to participate in various parts of the input sequence and capture diverse relationships. 

\subsubsection*{Self-Attention Mechanism}
For each word in the input, attention scores are calculated using the following components:
    \begin{itemize}
        \item {Query (Q)}: Represents the current word for which the attention score is being computed.
        \item {Key (K)}: Represents all words in the input sequence.
        \item {Value (V)}: Represents the actual values used to compute the weighted sum for the output.
    \end{itemize}    
The attention score for a pair of words is computed using the scaled dot product of the query and key, followed by a SoftMax function to obtain a probability distribution: 
    
$$
    \text{Attention}(Q, K, V) = \text{SoftMax}\left(\frac{QK^T}{\sqrt{d_k}}\right)V
$$
    
Here, $ d_k $ is the dimension of the key vectors, and the division by $ \sqrt{d_k} $ is a scaling factor to ensure stable gradients. After that, the attention scores compute a weighted sum of the value vectors, resulting in the self-attention output.

\subsubsection*{Multi-Head Attention}

To capture different types of relationships and dependencies, transformers use multi-head attention. This involves running multiple self-attention operations in parallel (each with different parameter sets) and then concatenating their outputs. This allows the model to jointly attend to information from different representation subspaces:
$$
\text{MultiHead}(Q, K, V) = \text{Concat}(\text{Attn}_1, \text{Attn}_2, \ldots, \text{Attn}_n)W_O
$$

\subsubsection*{Feed-Forward Networks (FFN)}

After the attention mechanism, the output is passed through a feed-forward neural network (FFN). This network consists of multiple linear transformations with non-linear activations in between:

$$
\text{FFN}(x) = \sigma(xW_1 + b_1)W_2 + b_2
$$

Here, $ W_1 $, $ W_2 $, $ b_1 $, and $ b_2 $ are learned parameters, and $\sigma$ is the activation function. The FFN is applied independently to each position in the sequence, allowing the model to learn complex representations.

\subsubsection*{Residual Connections and Layer Normalization}

Each sub-layer (attention and FFN) in the transformer is wrapped with residual connections \cite{resnet} and followed by layer normalization \cite{layernorm}. Residual connections help train deeper networks by allowing gradients to flow through the network directly. Layer normalization ensures that the input to each sub-layer has a stable distribution, which helps in faster convergence during training:
$$
\text{LayerNorm}(x) = \frac{x - \mu}{\sqrt{\sigma^2 + \epsilon}} \cdot \gamma + \beta
$$
where $ \mu $ and $ \sigma^2 $ are the mean and variance of $ x $, and $ \gamma $ and $ \beta $ are learned scale and shift parameters.

\subsection{Scope of the Survey on Large Language Models}
Based on previous research categorizations~\cite{yang2024harnessing}, we classify language models into three main types: Encoder-Decoder, Encoder-only, and Decoder-only models. 
All these models are based on the Transformer architecture~\cite{vaswani2017attention}. 
Encoder-decoder and Encoder-only models are considered BERT-style models~\cite{devlin2018bert}, while Decoder-only models are termed GPT-style models~\cite{gpt}.

The term ``large language model" lacks a precise definition and scope, leading to ongoing discussions in the field. 
For instance, Yang et al.~\cite{yang2024harnessing} consider the BERT model as a ``large" language model, yet their focus is predominantly on GPT-style models. 
Conversely, Zhao et al.~\cite{zhao2023survey} define BERT-style models as ``small-scale language models."
 
This survey focuses on GPT-style models, particularly those with model sizes equal to or larger than GPT-2~\cite{gpt-2}, which contains 1.5 billion parameters. This focus is based on three primary reasons:
\begin{itemize}
    \item {Shift in Popularity}: BERT models, especially Encoder-only variants, have gradually begun to fade away within the community~\cite{yang2024harnessing}. The landscape changed significantly after 2021 with the introduction of transformative models like GPT-3~\cite{gpt-3}, which led to a surge in the adoption of decoder-only architectures. GPT-style models have since dominated the development of LLMs.

    \item {Scaling Laws}: Extensive research demonstrates that increasing model size substantially enhances LLM capabilities. In 2020, OpenAI's introduction of the ``scaling law"~\cite{scaling_law} highlighted that model performance is strongly correlated with size. For example, GPT-3, with its 175 billion parameters, vastly outperforms BERT's 300 million parameters. The emphasis on ``large" models is a defining characteristic of GPT-style models, resulting in significantly different hardware and software solutions compared to BERT-style models.

    \item {Autoregressive Mechanism}: GPT-style models employ an autoregressive mechanism, which has proven superior in few-shot and zero-shot scenarios. However, this mechanism also introduces significant hardware performance challenges, which will be discussed in Section~\ref{sec:inferece}.
\end{itemize}

The advent of LLMs has revolutionized natural language processing and artificial intelligence. However, these models come with significant challenges, particularly in terms of computational and memory requirements, making efficient deployment a critical concern. Strategies are proposed to address these challenges in the training and inference phases from both the software and hardware perspectives.
This survey will focus on recent advancements in GPT-style models from aspects of the system, algorithm, and accelerator.

\section{Training}
\label{sec:train}
Training LLMs is a vital but both time- and resource-consuming step in their development. LLM training can be classified into two categories: 1) pretraining and 2) fine-tuning. 
Pretraining requires large datasets, many steps, and large batch sizes, making it very expensive. 
As reported in the literature~\cite{lepikhin2020gshard}, training a 600B model can take over 196,000 TPUv3 core hours. 
More optimized models require a much higher training cost. 
According to the study~\cite{llama2}, with NVIDIA 80GB A100 GPU under 400W power consumption, it takes over 184,000 GPU hours for pretraining Llama2-7B and over 1,720,000 hours for Llama2-70B. 
The electricity cost alone for training all four variants of Llama2 amounts to approximately \$158,000. 
Fine-tuning, on the other hand, can be performed with smaller datasets, fewer steps, and smaller batch sizes. The focus of this work is on the expensive pretraining step which will henceforth be referenced to as simply training.

At the scale of the LLM model size, both compute time and energy consumption per step are significant. Even marginal improvements in these areas could lead to substantial cost savings and reduced environmental impacts.
To improve training performance, it is critical to optimize various types of parallelism. 
Data parallelism is still effective. However, as the model size scales and the system becomes more distributed, it becomes increasingly difficult to eke out performance gains from data parallelism alone.
In addition, the peak performance of the hardware can limit the achievable data parallelism. 
To reduce energy consumption, the proposed framework must minimize data movement, and the supported hardware should be energy efficient. 
Coupled with performance and energy consumption challenges, LLMs have more stringent hardware requirements. 
First, it requires a large memory. Unlike inference, where only parameters will be stored, training needs parameters, gradients, optimizer states, and activations to be stored. 
Table \ref{tab:training_data_size} illustrates the relative size for each variable. 
Second, and correspondingly, LLMs require a higher memory bandwidth. 
As the size of the model increases, data movement becomes more intensive, leading to the need for high-bandwidth communication. This effect is even more pronounced when the system becomes distributed or when offloading techniques are applied, both of which incur increased data swapping.

\begin{table}[t]
\caption{Size of each type of data involved in training for a model with $N$ parameters and batch size as $B$. $x$ is a variable that depends on the model architecture. 
}
\label{tab:training_data_size}
\centering
\begin{tabular}{c c c c}
\hline
\textbf{Parameters} & \textbf{Gradients} & \textbf{Optimizer States}  & \textbf{Activations} \\
\hline
$2N$  & $0 \sim 2N$ & $4N \sim 12N$ &  $xNB$ \\
\hline
\end{tabular}
\end{table}

To address these challenges, academics and industry have proposed many solutions, ranging from infrastructure to hardware and algorithms. 
In particular, collaboration between hardware and software design is critical to addressing these challenges. In the following subsections, we will discuss solutions at each level and the challenges they target to address, including system, algorithm, and accelerator.

\subsection{Framework and System}
In this subsection, we start by introducing different types of parallelism and popular distributed infrastructures. Then, offloading techniques, a powerful solution addressing the issue of not enough memory, will be discussed. 
Furthermore, rematerialization and LoRA will be illustrated. Finally, we will introduce existing popular frameworks for training. 

\textbf{Parallelism.}
With the increasing complexity of DNN models, distributed training has become essential, especially for LLMs. An example of data parallelism in this domain is illustrated by the PyTorch Distributed Data-Parallel (DDP)~\cite{li2020pytorch} feature. 
DDP duplicates the setup to process different data portions simultaneously and synchronizes after each training step. 
Model parallelism splits the model across multiple GPUs, with each GPU handling different stages.
Model parallelism includes two categories: pipeline parallelism~\cite{lepikhin2020gshard,shoeybi2019megatron,xu2023efficient}, which assigns individual layers to single GPUs, and tensor parallelism~\cite{harlap2018pipedream,huang2019gpipe,johnstone1998memory}, which divides each tensor into chunks allocated to specific GPUs. 
In addition to traditional data and model parallelism, an emerging parallelism called fully sharded data parallelism (FSDP) is proposed in~\cite{zero} for LLM training.

\textbf{Memory Optimization.} 
Zero Redundancy Optimizer (ZeRO) \cite{zero} and its subsequent works \cite{zero-offload,zero-infinity,zero++} have been proposed to alleviate the high GPU memory requirement in large model training. ZeRO focuses on reducing redundant copies of data on GPU. It proposes three main optimization stages that partition the optimizer states, gradients, and parameters accordingly. 
ZeRO-Offload \cite{zero-offload} enables the training of even larger models by offloading optimizer states and optimizer updates to the CPU in order to strike a balance between accessibility and efficiency. 
ZeRO-Infinity \cite{zero-infinity} recognizes the much higher growth speed of model size than GPU memory and thus explores more methods that trade efficiency to enable the training of extremely large models. 
In addition to previous work, it incorporates NVMe memory for offloading for more storage space and offloads parameters, gradients, and activation checkpoints since their sizes can no longer be held on GPU as the model size grows. 
In addition, it manages the operations to reduce the buffer size requirement further. 
ZeRO++ \cite{zero++} returns to the design of ZeRO and focuses more on communication efficiency for large clusters of GPUs. \rev{While offloading can facilitate the training of large models, it significantly increases communication overhead between the GPU and CPU. This is because the connection between these components, such as PCIe in most system setups, typically offers limited bandwidth compared to the GPU's peak performance and memory bandwidth. As a result, this bottleneck can lead to substantial slowdowns during training.}

Rematerialization, also known as checkpointing or recomputation~\cite{chen2016training}, is a technique used in training LLMs to manage memory usage more efficiently by trading off computational resources. 
During the forward pass, only a subset of intermediate activations is stored, with the selection based on a strategy that minimizes memory usage while maintaining manageable computational overhead. 
During the backward pass, the necessary intermediate activations that were not stored are recomputed on the fly from the previously stored activations. Combining recomputed activations with stored activations enables the gradient calculation necessary to update the model parameters. 
Rematerialization significantly reduces memory usage, allowing for training larger models or using larger batch sizes without exceeding hardware memory limits. The primary trade-off is the increased computational cost due to recomputation, which is often acceptable given the benefits of training more complex models. 
\rev{This approach is akin to register allocation via graph coloring~\cite{chaitinacchm81}, which seeks scheduling strategies to maximize the reuse of limited registers.}
Rematerialization has been implemented in PyTorch for homogeneous sequential networks~\cite{chen2016training}, and more advanced versions~\cite{herrmann2019optimal} are modified for heterogeneous networks.

However, these memory reduction techniques, like recomputation and ZeRO, cause severe memory fragmentation. 
To address this, GMLake~\cite{guo2024gmlake} employs a virtual memory stitching (VMS) mechanism to merge non-contiguous memory blocks, significantly reducing GPU memory usage for LLM fine-tuning. 
Transparent to DNN models and techniques, GMLake ensures the seamless execution of resource-intensive tasks.

\textbf{Popular Frameworks.} Besides being able to be implemented in the conventional PyTorch~\cite{li2020pytorch} and Tensorflow~\cite{abadi2016tensorflow}, there exist emerging frameworks which are specialized for LLM training like DeepSpeed~\cite{deepspeed}, Hugging Face Transformer~\cite{wolf2019huggingface}, Torchtune~\cite{torchtune}, Megatron-LM, etc.
Most frameworks are integrated with the optimization techniques for LLM training mentioned above.

\subsection{Algorithm and System Co-design}
Full fine-tuning (fine-tuning all learnable parameters) of a pre-trained LLM for a specific downstream task is often infeasible or too costly. 
Full fine-tuning poses non-trivial challenges to the supporting system platforms due to its computationally-intensive and resource-demanding nature and can potentially hurt the generalizability of the pre-trained backbone model. 
Parameter Efficient Fine-Tuning, or PEFT, addresses this need by either introducing additional lightweight trainable modules or selectively adapting a small fraction of the original parameters.
The family of adapter-based PEFT methods inserts extra trainable parameters strategically either within the frozen pre-trained transformer blocks or as attached components. 

\textbf{LoRA.} LoRA, or Low-Rank Adaptation, is a technique designed to fine-tune large pre-trained models in a parameter-efficient manner~\cite{hu2021lora}. 
Instead of updating all model parameters during adaptation, LoRA focuses on a lower-dimensional subspace by applying a low-rank decomposition to the weight matrices. 
This involves updating pairs of smaller matrices that can approximate the necessary changes, significantly reducing the number of parameters to be fine-tuned. 
This approach makes the fine-tuning process faster and less memory intensive, which is particularly advantageous for large models. 
Despite reducing parameters, LoRA achieves competitive performance with traditional fine-tuning methods, making it an attractive option for adapting large pre-trained models to specific tasks without high computational costs.

Due to its simplicity and effectiveness, many LoRA-variant have been proposed to improve upon the vanilla LoRA. 
SPLoRA~\cite{hedegaard2022structured} and LoRAPrune~\cite{zhang2024loraprunepruningmeetslowrank} both leverage structured channel-pruning to remove groups of weights in order to increase the computational efficiency. 
QLoRA~\cite{dettmers2023qlora}, a highly memory-efficient technique that first quantizes a pre-trained model into a novel 4-bit NormalFloat data type with an innovative method called Double Quantization and then backpropagate the gradients through the quantized weights with Paged Optimizers, can finetune a LLaMA 65B parameter model on a single 48GB GPU with similar performance as that of a 16-bit full-finetuned model. 
LQ-LoRA~\cite{guo2024lqlora} further extends the quantization limit to sub-3 bits by decomposing each pre-trained matrix into a frozen quantized matrix and an adaptable low-rank matrix. 
QA-LoRA~\cite{xu2024qalora} tries to get the best of both quantization and adaptation by balancing the unequal freedom between them inherent in LoRA through group-wise operators.

\rev{\textbf{Prompt-based Learning.} Prompt-based learning has become increasingly prominent in the field of large language models (LLMs), primarily by leveraging minimal examples or specific cues to steer a pre-trained language model (PLM) toward generating the desired output. This approach marks a departure from traditional supervised learning, which relies on extensive labeled data to train a model explicitly. The advent of OpenAI's GPT-3~\cite{brown2020language} significantly advanced the exploration of prompt-based learning, demonstrating that the massive scale of GPT-3 enables the generation of relevant outputs with well-designed prompts without necessitating task-specific model fine-tuning. Despite this, manually crafted prompts often exhibit a performance discrepancy compared to fine-tuned models, as noted by multiple studies~\cite{brown2020language,schick2020exploiting,gao2020making,sun2022paradigm}. Recent advancements have shown that prompts need not be confined to natural language forms but can also be optimized in a continuous space using gradient descent, enhancing their efficiency~\cite{li2021prefix,hambardzumyan2021warp,qin2021learning,liu2023gpt,zhong2021factual,liu2021p}. In scenarios where only the continuous prompts are tuned—keeping the PLM's parameters unchanged—the training remains efficient while achieving comparable performance to full model tuning. The concept of prompt tuning~\cite{lester2021power,li2021prefix,sun2023fedbpt} was introduced to fine-tune a continuous vector that is concatenated to the input embeddings, optimizing the prompt directly within the embedding space. Building on this, the p-tuning methodology was developed to further enhance performance by learning concrete prompts within the embedding space, a technique further refined in subsequent studies~\cite{liu2021p,liu2022p,liu2023gpt}.}

\textbf{Retrieval-Augmented Generation.} Retrieval-augmented generation (RAG) is a technique that enhances the capabilities of generative models (like large language models) by integrating external knowledge retrieval systems. RAG is a powerful technique for enhancing LLMs, allowing them to generate responses that are grounded in up-to-date, accurate information. By combining the strengths of retrieval systems and generative models, RAG ensures that large language models are more reliable, less prone to hallucination, and better suited for real-world applications that demand accuracy and specificity. Currently, approximate nearest neighbor search (ANNS), which retrieves the approximate nearest neighbors of a given query in the high-dimensional and large-scale vector database, has been widely used as an RAG technique. 
Hierarchical Navigable Small World~\cite{hnsw} (HNSW) is a graph-based ANNS algorithm that organizes data points in multiple layers of proximity graphs, enabling fast searches by navigating through a hierarchical structure. DiskANN~\cite{jayaram2019diskann}, on the other hand, is designed for handling large datasets that don't fit into memory by extending nearest neighbor search to disk-based systems, offering a balance between speed and memory efficiency. Faiss~\cite{faiss} is a popular library developed by Meta, providing highly optimized algorithms for similarity search, especially leveraging GPU acceleration for fast nearest-neighbor computations. These ANNS algorithms make it feasible to efficiently handle the retrieval tasks requiring high-dimensional similarity search in real-time.

\textbf{Others.}
Other PEFT methods selectively update a subset of the pre-trained model weights during adaptation. 
Diff pruning\cite{guo-etal-2021-parameter} aims to learn an additive sparse mask that is applied to the frozen pre-trained weights for each task, effectively localizing task-specific weights to update. 
PaFi~\cite{liao-etal-2023-parameter}, on the other hand, finds a universal set of parameters to update for all downstream tasks based on parameter magnitude. 
FISH Mask~\cite{sung2021training} gauges the importance of each model parameter by estimating its' Fisher information, resulting in a binary mask that indicates which parameters are crucial for the current task.

\subsection{Accelerators for Training}
LLM training and fine-tuning require substantial computational resources. Traditional CPUs, while versatile, are often insufficient for the massive parallel processing demands of LLMs. 
This has led to the adoption and innovation of specialized hardware accelerators designed to enhance the efficiency and speed of training and fine-tuning processes.
For LLM, the extremely large volume of memory footprint makes the accelerator mainly focus on memory-centric optimizations, especially for memory compression.

\textbf{GPU.}
GPUs (Graphics Processing Units) have become the cornerstone of modern deep learning infrastructure. Originally designed for rendering graphics, their highly parallel architecture makes them ideal for the matrix and tensor operations fundamental to training neural networks. GPUs, such as NVIDIA’s A100~\cite{nvidia-a100} and H100~\cite{nvidia-h100}, offer significant speedups in training times compared to CPUs, making them a popular choice for both academic research and industrial applications.
For LLM, NVIDIA proposed a specific optimization for the training process, called Transformer Engine~\cite{nvidia-h100}, based on the mix-precision training technology~\cite{micikevicius2018mixed}.

\textbf{Transformer Engine.}
The NVIDIA Transformer Engine is designed to optimize the performance and efficiency of transformer-based models widely used in natural language processing and AI. 
It leverages mixed-precision techniques, combining FP16 (16-bit floating point) and FP32 (32-bit floating point) computations to maximize throughput and minimize memory usage without compromising model accuracy. Using Tensor Cores on NVIDIA GPUs, the Transformer Engine accelerates training and inference processes, enabling faster development and deployment of AI applications. This approach enhances computational efficiency and reduces costs and energy consumption in large-scale AI operations.

\textbf{TPU.}
Tensor Processing Units (TPUs) are custom-built accelerators developed by Google specifically for machine learning tasks, particularly deep learning and large language models (LLMs). TPUs are designed to handle the vast computational requirements of training LLMs by providing high throughput and efficient performance. They utilize a systolic array architecture~\cite{systolic} to accelerate matrix multiplications, which are fundamental to neural network operations. TPUs support training and inference, with features such as high memory bandwidth and mixed-precision computation to enhance speed and efficiency. TPUs significantly reduce the time and cost of training large, complex models by offering scalable, high-performance computing.

\textbf{Others.}
ASIC accelerator designs also aim to reduce the memory bottleneck in large language models (LLMs). Smart-Infinity~\cite{smart-infinity} is the first study to leverage host memory and storage as an extended memory hierarchy for LLM training, enhancing efficiency by addressing storage bandwidth bottlenecks through near-storage processing. It performs parameter updates on custom near-storage accelerators, significantly reducing storage traffic. The system includes an efficient data transfer handler to manage system integration and overlap data transfers with fixed memory consumption. Additionally, accelerator-assisted gradient compression/decompression improves scalability by reducing write traffic. 
Before this, many studies focused on efficient training based on sparsity and quantization, such as Sigma~\cite{qin2020sigma}, TensorDash~\cite{tensordash}, FAST~\cite{zhang2022fast}, and Combricon-Q~\cite{zhao2021cambricon}, including evaluations on Transformer-based models. However, these studies mainly targeted much smaller language models, like GNMT~\cite{gnmt}.

\rev{
Efficient training for large language models (LLMs) is a promising yet nascent field. Despite its potential, practical challenges and deployment difficulties have limited research, particularly in accelerator design. In the next section, our survey focus will shift to inference optimization studies, which present lower complexity and broader applicability.
}
\section{Inference}
\label{sec:inferece}
LLMs are powerful and capable models, but deploying pre-trained LLMs is often difficult due to the models' exceptionally high resource usage requirements. In extreme cases, the largest models, such as LLaMa-70B, contain tens of billions of parameters, incurring massive computational and memory costs impractical for most consumer-grade devices. As such, much research has been performed to mitigate these bottlenecks, such as using dedicated accelerators, advanced model compression methods, and algorithmic advances. The following subsections offer discussions and key insights into these solutions.

\subsection{LLM Inference System}
A critical step for LLMs is their deployment on hardware devices, catering to both offline inference and online serving scenarios. Offline inference involves a single user with all requests initiated at the start, aiming to reduce inference latency by enhancing the model's forward process. 
In contrast, online serving handles asynchronous requests from multiple users, requiring optimized memory management and efficient batching and scheduling strategies to improve throughput. 
In addition, the increasing scale of LLMs generally necessitates deployment across multiple hardware devices, creating an intricate infrastructure. 
Consequently, system-level optimization has become a significant research focus. This section explores key optimization techniques.

\subsubsection{Inference Engine}
Inference engine optimizations for LLMs aim to accelerate the forward process, achieved through both fusion and non-fusion based techniques.

\textbf{Operation fusion.}
Kernel fusion is a widely adopted optimization technique for LLM inference. It involves combining multiple operators or layers in the computation graph.
This method enhances computational efficiency by reducing memory access, decreasing kernel launch overhead, and improving parallelism without data dependencies. 
Profile results indicate that attention and linear operations dominate LLM runtime, accounting for over 75\% of total inference duration \cite{profile_time}. 
To optimize attention computation, FlashAttention~\cite{flashattention, flashattention2} integrates the entire process into a single operator, achieving 3$\times$ training speed up on the GPT-2 model.
FlashDecoding~\cite{flashdecoding} and FlashDecoding++~\cite{flashdecoding_plusplus} further enhance this by optimizing parallelism and introducing efficiency in SoftMax computation. 
For linear operations, TensorRT-LLM \cite{tensorrt-llm} employs a specialized GEMV implementation, while FlashDecoding++ \cite{flashdecoding_plusplus} adapts FlatGEMM for reduced dimensions, utilizing fine-grained tiling and double buffering to improve efficiency. 
Additional optimizations include the fusion of lightweight operations such as LayerNorm, SwiGLU, activation functions, and residual additions by frameworks like DeepSpeed~\cite{deepspeed}, ByteTransformer~\cite{bytetransformer}, xFormers~\cite{xFormers}, and TensorRT-LLM~\cite{tensorrt-llm}, which also uses a pattern-matching algorithm to identify potential fusions across various LLM architectures.

\textbf{Memory Optimization.}
Beyond fusion, addressing the challenges posed by the dynamic sizes of input and output tokens during inference is crucial. Inspired by CPU virtual memory systems, vLLM~\cite{vllm} introduces PagedAttention to segment the KV cache into manageable blocks, enhancing memory management with 24$\times$ higher throughput than HuggingFace Transformers \cite{wolf2019huggingface}. 
When GPU memory is insufficient, techniques such as ZeRO-Inference by DeepSpeed~\cite{deepspeed} offload large model weights to CPU memory to improve performance by overlapping computation with weight fetching. 
Similarly, FlexGen~\cite{flexgen} employs a linear programming-based strategy to optimize offloading across CPU, GPU, and disk spaces. The utilization of high-capacity flash memory for storing model parameters further demonstrates efficient inference by optimizing memory usage \cite{llminflash}.

\subsubsection{Online Serving}
Optimizations in LLM serving systems are centered around effectively managing asynchronous requests to boost both throughput and responsiveness, utilizing strategies in dynamic batching, memory management, and Scheduling.

\textbf{Batching Optimization.}
Efficient handling of variable request sizes is a primary concern in LLM serving. 
ORCA~\cite{Orca} introduces continuous batching or rolling batching, where new requests are dynamically batched as previous ones complete, optimizing the use of computational resources. 
This method is extended in Sarathi~\cite{sarathi}, Sarathi-Serve~\cite{sarathiserve} and LightLLM~\cite{lightllm}, which employ a split-and-fuse technique to balance load across different processing stages, thereby minimizing response times and enhancing throughput.

\textbf{Memory Management.}
Efficient memory usage is also crucial due to the extensive requirements of the KV cache, especially for lengthy context interactions. Traditional allocation strategies often lead to substantial memory waste. To address this, S$^3$~\cite{jin2023s} predicts the upper limit of generation lengths, optimizing the initial memory allocation. 
Further improvements are seen in vLLM~\cite{vllm}, which introduces a paged storage mechanism similar to that used in operating systems, allocating the largest possible contiguous space and mapping KV caches dynamically to reduce fragmentation. LightLLM~\cite{lightllm} refines this approach by allocating KV cache storage at the token level, maximizing space utilization, and minimizing waste.
LLM in a Flash \cite{llminflash} addresses the challenge of efficiently running large language models (LLMs) that exceed the available DRAM capacity by storing the model parameters in flash memory and dynamically loading them into DRAM as needed. When a new token is added, the system only needs to update a minimal number of neurons rather than reloading all neurons.

\textbf{Scheduling Strategy.}
Variability in request length can significantly impact scheduling efficiency. Traditional first-come-first-served approaches often lead to inefficient resource utilization, known as head-of-line blocking \cite{vllm, Orca, lightllm}. 
To combat this, FastServe~\cite{fastserve} leverages a preemptive scheduling strategy that prioritizes requests based on their estimated completion time, thus improving throughput and reducing job completion times. 
Additionally, VTC~\cite{vtc} introduces a fairness-oriented scheduling model that adjusts resource allocation based on the workload of incoming requests, ensuring a balanced service across different users. 
Scheduling in Distributed architectures offers unique opportunities for scaling LLM services. SpotServe~\cite{spotserve} addresses the challenges of using cloud-based preemptible GPU resources by implementing strategies for dynamic adjustment and state recovery, ensuring robust service continuity.
Finally, techniques like those proposed in Splitwise~\cite{splitwise} and TetriInfer~\cite{tetriinfer} disaggregate compute-intensive prefilling from memory-intensive decoding processes, tailoring resource allocation to the specific demands of each stage.

\textbf{Heterogeneous Computing.}
The significant computational and memory demands of large language model (LLM) inference typically require multiple high-end accelerators. 
However, driven by the growing need for latency-insensitive tasks, some studies~\cite{llminflash, xue2024powerinfer, yin2024llm, sheng2023flexgen, song2023powerinfer} explore high-throughput LLM inference using limited resources, such as a single GPU, edge devices, and mobile devices.
The most critical challenge is data transfer due to insufficient memory capacity. 
There are typically two scenarios of data transfer: the first \cite{llminflash, xue2024powerinfer, yin2024llm} is when model parameters and intermediate results need to be stored in storage (e.g., Flash memory) due to limited DRAM capacity, resulting in data transfer between DRAM and storage; 
the second scenario occurs when CPU and GPU cannot share memory \cite{sheng2023flexgen, song2023powerinfer}, requiring model parameters and intermediate results to be stored in host memory due to limited GPU memory, thus leading to data transfer between CPU and GPU. 
Reducing the cost of data transfer often becomes a primary consideration for optimizing LLMs on edge devices.

These studies can optimize the system from multiple angles to reduce data transfer and lower storage costs. 
Existing work has observed that retaining only a subset of effective activation values does not degrade model performance, and the sparsity pattern of activation values is predictable \cite{liu2023deja, song2023powerinfer, xue2024powerinfer, llminflash, song2024turbo}. By leveraging the sparsity of activation values, only a subset of model parameters is needed for computation, significantly reducing data transfer and storage costs.

This section effectively delineates the optimization strategies for deploying large language models (LLMs) in both offline and online contexts, focusing on enhancing system performance through various techniques such as operation fusion, memory optimization, and dynamic batching. The outlined approaches, from kernel fusion like FlashAttention~\cite{flashattention} to memory-efficient strategies like vLLM~\cite{vllm} and scheduling optimizations such as FastServe~\cite{fastserve}, reveal the depth of innovation aimed at improving the responsiveness and throughput of LLM systems. However, the practical implementation of these techniques often involves trade-offs between computational efficiency, memory usage, and response times. Real-world performance data would be invaluable in quantifying these trade-offs, offering a clearer perspective on the effectiveness of different strategies in varied deployment scenarios. Such data could guide in selecting the most appropriate optimizations based on specific requirements, such as latency constraints or hardware limitations, ensuring optimal performance tailored to the needs of diverse models or applications.
\subsection{Algorithm for Efficient LLM}
Faster inference is essential for large models, especially those with commercial potential. Different algorithms tailored for various aspects of large models have been proposed to improve the inference efficiency. 
In this subsection, we introduce several techniques that have greatly impacted the community.
Specifically,  \textit{Mixture-of-Experts (MoE)} speeds up the feed-forward networks (FFNs), \textit{Efficient Attention} speeds up the attention module, \textit{Speculative Decoding} allows faster auto-regression and \textit{Structured State Space Models (SSMs)} serve as an alternative to transformers that improve the computational efficiency over long sequences. 

\subsubsection{MoE}
Mixture-of-experts (MoE) was first proposed in \cite{jordan1994hierarchical, Jacobsadaptive} by Michael I. Jordan and Robert A. Jacobs more than three decades ago. 
The initial insight was to propose an architecture such that each expert handles a different subset of input data. 
Later on, \cite{eigen2013learning} proposes to stack several neural-network-based MoE layers to make the model deeper with the rise of deep learning. 
Recently, the era of large models came under the guidance of the scaling law~\cite{kaplan2020scaling}, asserting that the performance of the model demonstrates a predictive behavior as the model size increases. 
Now, MoE attracts more and more attention due to its scalability; that is, drastically increasing the number of parameters incurs little computational overhead. MoE offers a solution for fast inference of large models and has been employed by several famous LLMs, such as GPT-4~\cite{achiam2023gpt} and Mixtral~\cite{jiang2024mixtral}.

An MoE layer consists of several expert models and a router function (or gating function in some literature). 
The router function will select experts for computation for each input entity. Specifically, let $g(\cdot)$ denote the router function and $\{f_i(\cdot)\}_{i=1}^E$ denote $E$ experts. The output of an MoE layer is as follows: 
\begin{align}
    \label{eq:moe}
    M(\bm{x})  &= \sum_{i\in\mathcal{I}} p_i(\bm{x})f_i(\bm{x}),\\ \text{where  }
    p_i(\bm{x})&=\frac{\exp\{h_i(\bm{x})\}}{\sum_{j=1}^E \exp\{h_j(\bm{x})\}}.
\end{align}
Let \( M \) denote the Mixture of Experts (MoE) model and \( h \) the router function. Typically, \( h \) is a linear model that performs a linear classification over experts. We use \( p \) to denote the probability distribution over all experts. The set \( \mathcal{I} \) signifies the selected experts; different choices for \( \mathcal{I} \) yield various MoE techniques.

Many breakthroughs have been made in large language models (LLMs) using Mixture of Experts (MoE). Shazeer et al. ~\cite{shazeer2017outrageously} developed an LSTM model with 137B parameters, significantly enhancing model capacity with minimal computational overhead. 
Switch Transformer ~\cite{fedus2022switch} extended this to a transformer-based MoE model with 1.6T parameters, confirming MoE expansion follows the scaling law. GShard ~\cite{lepikhin2020gshard} efficiently implements large-scale MoE, expanding it by over 600B parameters. Clark et al. ~\cite{clark2022unified} investigated the scaling law for routing-based language models, deriving an Effective Parameter Count for scalable models.

Addressing instability in training large MoE models, ST-MoE ~\cite{zoph2022st} improved transfer learning performance. 
Mixture-of-Depth (MoD) ~\cite{raposo2024mixture} explores skipping layers in transformer models. Xue ~\cite{xue2022go} and Riquelme ~\cite{riquelme2021scaling} develop vision transformer-based MoE models, with Riquelme achieving state-of-the-art performance with half the computation. Obando ~\cite{obando2024mixtures} constructs an MoE model for reinforcement learning, providing empirical evidence for scaling laws in this domain.
MoEfication ~\cite{zhang2021moefication, zhu2024llama} converts dense models to MoE models by grouping weights in FFNs. Routing strategies have been explored to balance load and address training instability. Base Layer ~\cite{lewis2021base} uses a linear assignment problem for balanced loading, while Hash Layer ~\cite{roller2021hash} uses predefined hash functions. StableMoE ~\cite{dai2022stablemoe} stabilizes training by learning a balanced router function. Differentiable MoE architectures like Soft MoE ~\cite{puigcerver2023sparse} and Lory ~\cite{zhong2024lory} enhance stability by making operations differentiable.
MoE models, which activate only some weights during each forward pass, improve GPU memory efficiency and throughput. SE-MoE ~\cite{shen2022se}, M3ViT ~\cite{fan2022m3vit}, and SiDA-MoE ~\cite{du2024sida} introduce strategies to optimize throughput and expert caching. The analogy to MoE, FoE (Fusion of Experts)~\cite{wang2024fusing} also aggregates knowledge from different experts, which can be pre-trained in their respective domains.

\subsubsection{Efficient Attention}
The bottleneck of transformers on computation is the attention scheme both in time and memory. 
Efforts have been made on algorithms towards efficient attention schemes that either speed up the inference or improve the memory usage. 
The main lines of research can be split into two categories, one is grouping the keys and values and the other is approximating the attention score either by kernel methods or low-rank methods. 

\textbf{Multi-Query Attention.} 
Multi-Query Attention (MQA)~\cite{shazeer2019fast} and Group-Query Attention (GQA)~\cite{ainslie2023gqa} improve the attention schemes by sharing the keys and values in multi-head attention. 
Specifically, in multi-head attention, each head possesses a pair of keys and values. MQA averages all the keys and values across all heads and shares the averaged keys and values for all heads. 
The proposed attention scheme significantly saves the memory bandwidth for loading keys and values and speeds up the decoding process. 
However, MQA may lead to performance degradation. GQA builds upon MQA and tackles the quality degradation by relaxing the sharing across heads. GQA defines a hyperparameter $G$ that denotes the number of groups where, within each group, keys and values are averaged and shared. 
For example, GQA-$1$ reduces to MQA, and GQA-$H$ reduces to multi-head attention with $H$ heads.

\textbf{Attention Approximation.} Attention Approximation techniques improve the efficiency of attention schemes by reducing the computation of the attention matrix from $O(n^2)$ to $O(n)$, where $n$ is the sequence length. 
\textit{Kernel-based methods} aim to design a kernel feature map $\phi\in\mathbb{R}^{n\times d}$, where $d$ is the feature dimension. The formulation of kernel-based methods is as follows: 
\begin{align}
    \text{SoftMax}(QK^T)V \approx \phi(Q)\phi(K)^TV,
\end{align}
where $\phi(K)^TV$ is a multiplication between $\mathbb{R}^{d\times n}$ and  $\mathbb{R}^{n\times d}$ and $\phi(Q)(\phi(K)^TV)$ is a multiplication between $\mathbb{R}^{n\times d}$ and $\mathbb{R}^{d\times d}$, taking $O(nd^2)$ complexity. 
Performers~\cite{choromanski2020rethinking} and RFA~\cite{peng2021random} employ the random feature projection as the feature map, while PolySketchFormer~\cite{kacham2023polysketchformer} exploits sketching techniques with polynomial functions. 
\textit{Low-rank-based methods} aim to use low-rank matrix compression techniques to change $Q\in\mathbb{R}^{n\times d}$ and $K\in\mathbb{R}^{n\times d}$ to $\Tilde{Q}\in\mathbb{R}^{k\times d}$ and $\Tilde{K}\in\mathbb{R}^{k\times d}$, where $k$ is a smaller number. 
Thus, the computational complexity for low-rank-based methods is $O(nk^2)$. Linformer~\cite{wang2020linformer} is the first to explore the possibility of low-rank approximation of the attention matrix. 
LRT~\cite{zhang2023lrt} then proposes to apply low-rank approximation on both the attention matrix and feed-forward layers. 
FLuRKA~\cite{gupta2023flurka} combines kernel-based methods and low-rank-based methods that first apply low-rank approximation and then apply kernel feature map on the low-rank $\Tilde{Q}$ and $\Tilde{K}$. 

\textbf{Speculative Decoding.}
Speculative decoding~\cite{leviathan2023fast, kim2024speculative} aims to speed up the decoding process for auto-regressive large language models (LLMs). 
The motivation comes from the observation that memory loading is a bottleneck in LLM inference, and models of smaller sizes can output the correct tokens while memory is efficient. 
Specifically, speculative decoding employs a small model to generate tokens, and the LLMs consistently evaluate the draft generated by the small model to decide whether to accept or reject the generation. Upon rejecting small models, a resampling from LLM will be performed.  
Inference with large language models is primarily constrained by heavy I/O, which acts as the bottleneck. Speculative decoding significantly reduces memory I/O during inference, leading to improvements in latency, though it potentially increases FLOPs. 

Research on speculative decoding focuses on improving the acceptance rate from LLMs over models' generation. One line of research focuses on exploiting the computing units that ask small models to generate several candidates to be evaluated by LLMs in parallel~\cite{sun2024spectr, miao2023specinfer, xu2023llmcad}. 
The other line of research aims to tackle the problem through the lens of algorithms, improving the alignment between LLMs and small models~\cite{zhou2023distillspec, liu2023online, zhang2023draft}. 

\textbf{SSMs.}
The State-Space Models (SSMs), which are efficient yet effective, serve as an alternative to the transformer.  
SSMs are especially good at long sequence tasks given their linear computation and memory compared to transformers. 
The key idea of SSMs is to compress the input sequence of length $L$, $\{h_t\in \mathbb{R}^{d_\text{emb}} \}_{t=1}^{t=L}$, to a sequence of states $\{x_t \in \mathbb{R}^{d_\text{states}}\}_{t=1}^{t=L}$ based on HiPPO theory~\cite{gu2020hippo}. Compared to transformers, where the attention score is computed between every two embeddings in the sequence, SSMs compress all the up to $t$ embedding vectors in the sequence to the state $x_t$ and performs the prediction only based on the state by the following formulas:
\begin{align}
    \label{eq:ssm}
    x_t &= Ax_{t-1} + Bh_t, \\
    y_t &= Cx_t,
\end{align}
where $x$ is the state, $h$ is the input sequence, and $A, B$ and $C$ represent the transition matrices. SSMs enjoy linear computation and memory since at each round of propagation, the next token interacts with the states only instead of all previous tokens. 

Based upon the foundational architecture, the mainstream research focuses on better parametrization on transition matrices~\cite{gu2021combining, gu2021efficiently, smith2022simplified, gu2023mamba} and better computational architecture based on SSMs~\cite{smith2022simplified, gu2023mamba, dao2024transformers, park2024can, lieber2024jamba}. 
Specifically, LSSL~\cite{gu2021combining} proposes to initialize matrix $A$ via the optimal transition matrix proposed in~\cite{gu2020hippo}. 
Further, LSSL trains an SSM model through telescoping propagation equations~\ref{eq:ssm}, which can be computed efficiently through the Fast Fourier Transform. S4~\cite{gu2021efficiently} employs a diagonalized transition matrix $A$ to enhance the computational efficiency. 
Later on, S5~\cite{smith2022simplified} proposes to share the transition matrices across all input dimensions to boost the computational efficiency, while Mamba~\cite{gu2023mamba} and S4~\cite{gu2021efficiently} propose input dependent transition matrices that improve the model capability. Meanwhile, Mamba~\cite{gu2023mamba} and S4~\cite{gu2021efficiently} utilize a parallel scan technique that improves the computational efficiency of SSMs.  
MambaFormer~\cite{park2024can} and Jamba~\cite{lieber2024jamba} improve the SSM architecture by combining transformers into SSMs, where  \cite{park2024can} use the SSM layer to replace the FFN layer in transformers and Jamba~\cite{lieber2024jamba} add four transformer layers to SSMs. 
Mamaba2~\cite{dao2024transformers} proposes a new architecture based on State-Space Duality that achieves 2-8$\times$ speed up compared to Mamba~\cite{gu2023mamba}.

\subsubsection{Multi-modal LLMs.}
Advancements in text-only LLMs have paved the way for the rapid development of \textit{multi-modal} LLMs \cite{liu2023llava}, which can process visual inputs such as images and videos. 
The construction of these multi-modal LLMs follows a well-established recipe. Initially, a vision encoder (e.g., CLIP \cite{clip}) is used to encode the visual input into a sequence of embeddings. 
These embeddings are then passed through a projector, such as a multi-layer perceptron (MLP), to align them with the embedding space of the originally text-only LLM. 
Once aligned, the vision embeddings are concatenated with the text embeddings in an autoregressive manner, enabling the LLM to process and generate outputs based on both modalities.
However, the integration of vision tokens/embeddings introduces a significant computational burden compared to text-only LLMs. 
This is primarily due to the large number of vision tokens—often numbering in the hundreds or thousands—required to represent the visual input comprehensively \cite{liu2023llava,liu2024llavanext}.

To mitigate the increased computational cost associated with multi-modal LLMs, several methods have been developed to prune the number of vision tokens. 
One such method involves the use of the Perceiver module \cite{perceiver}, which employs a transformer with queries to perform learned pooling, effectively replacing the MLP as the projector \cite{idefics2}. 
This approach can significantly reduce the computational demands of both training and inference. 
Another method is the algorithmic approach of PruMerge \cite{prumerge}, which selectively retains important vision tokens while merging the less significant ones.
Despite these advancements, there remains considerable potential for further development in this area, as systematic explorations are still relatively limited. 
We believe that continued research and innovation will yield even more efficient techniques for handling vision tokens in multi-modal LLMs.

\subsection{Compression Methods and Accelerators}
The immense size of LLMs creates significant challenges for deployment, both due to the computational complexity as well as resource availability requirements. Significant research has been performed to strategically compress LLMs in order to mitigate these bottlenecks while preserving the capabilities of the model, increasing inference efficiency while continuing to scale down the required resources needed to execute.
Model compression can be categorized into four main methods: quantization, pruning, knowledge distillation, and low-rank factorization.

\subsubsection{Quantization}
Quantization is a highly effective method for reducing the size and computational demands of deep neural network (DNN) models. There are two primary quantization techniques: quantization-aware training (QAT) and post-training quantization (PTQ). QAT, as discussed in ~\cite{jacob2018quantization, wang2019learning, zhuang2021effective}, involves retraining the model to adapt to quantization noise. On the other hand, PTQ~\rev{\cite{gupta2015deep, jacob2018quantization, guo2022squant}} converts a floating-point model to a lower-bit model without requiring training data, making it particularly suitable for large-scale language models.

\textbf{Quantization Algorithm.}
Innovative quantization methods have significantly enhanced the efficiency and performance of LLMs. SmoothQuant~\cite{xiao2023smoothquant} enables 8-bit weight and activation quantization (W8A8) by migrating quantization difficulty from activations to weights through per-channel scaling, reducing activation outliers and maintaining accuracy. 
AWQ (Activation-aware Weight Quantization)~\cite{lin2023awq} optimizes low-bit weight-only quantization by protecting critical weights based on activation distributions, preserving model performance without backpropagation. 
QuaRot~\cite{ashkboos2024quarot} achieves outlier-free 4-bit inference using randomized Hadamard transformations, efficiently handling activation quantization by removing outliers. 
QuIP~\cite{chee2024quip} facilitates 2-bit quantization through incoherence processing, leveraging incoherent weight and Hessian matrices, and using adaptive rounding to minimize quantization error, supported by theoretical analysis. 

\textbf{Quantization Accelerator.}
To improve the accuracy of quantized DNN models, numerous studies have proposed new architecture designs based on advanced quantization techniques.
BitFusion~\cite{sharma2018bit} supports various bit-width quantizations by combining low-bit processing elements. 
OLAccel~\cite{zadeh2020gobo} and GOBO~\cite{zadeh2020gobo} quantizes outliers with higher precision, but these approaches often suffer from unaligned memory accesses, leading to additional overhead and limited computing speed.
ANT~\cite{guo2022ant} offers a fixed-length adaptive quantization framework, considering tensor distribution but overlooking outliers' importance. 
Mokey~\cite{mokey} uses narrow fixed-point inference for transformer models by converting values to 4-bit indices into dictionaries of 16-bit fixed-point centroids, improving hardware efficiency without fine-tuning.
OliVe~\cite{olive} employs an outlier-aware quantization method using an outlier-victim pair mechanism to address quantization challenges, reducing hardware overhead and aligning memory access, enabling efficient 4-bit quantization for weights and activations. These methods collectively advance the deployment of quantized LLMs in resource-constrained environments by improving performance, reducing memory usage, and maintaining model accuracy.

\textbf{New Data Type.}
Many studies also focus on designing new numeric types with reduced precision to improve model compression and efficiency. 
Microsoft Floating Point (MSFP)~\cite{msfp} uses a shared exponent for groups of values, enabling efficient dot product computations and higher arithmetic density compared to formats like Bfloat16 or INT8, making it ideal for large-scale cloud deployments. 
FP6-LLM~\cite{xia2024fp6llmefficientlyservinglarge} introduces a 6-bit floating-point format that leverages TC-FPx, a GPU kernel design, to reduce inference costs and improve performance for large language models (LLMs). 
LLM-FP4~\cite{Liu_2023} utilizes 4-bit floating-point quantization, optimizing exponent bits and clipping ranges, achieving minimal accuracy loss while enabling efficient deployment in resource-constrained environments. 
LLM.int8()~\cite{NEURIPS2022_c3ba4962} enables efficient 8-bit matrix multiplication by combining vector-wise quantization and mixed-precision decomposition, maintaining accuracy for models up to 175B parameters and reducing memory usage, facilitating inference on large models using consumer-grade GPUs.

Quantization offers significant benefits for large language models, primarily by reducing memory bandwidth and capacity requirements, which in turn accelerates performance for the memory-bound decoding process. As context lengths grow, weight quantization alone becomes insufficient, necessitating the quantization of the KV cache to maintain efficiency. However, it's important to note that pushing quantization to extremely low bit widths may not always yield proportional improvements due to the increased overhead in decoding operations.

\subsubsection{Sparsity}
Sparsity, which involves setting parts of the weights or activations to zero, is a commonly used technique for compressing neural networks. 
By efficiently skipping these zeros during inference, sparsification reduces computational complexity, memory occupancy, and bandwidth requirements. 
In LLMs, sparsification is applied to weights in fully connected layers and activations in attention scores, leading to two main techniques: weight pruning and sparse attention.

\textbf{Weight Pruning.}
Weight pruning~\cite{han2015deep, albericio2016cnvlutin, zhang2016cambricon, zhou2018cambricon,  zhu2019sparse, guo2020accelerating,  wang2021dual,  guan2024fractal, guo2024accelerating} reduces the number of parameters by removing less important weights. 
This process identifies and zeros out weights that have minimal impact on the model's performance, effectively compressing the model and making it more efficient.
To leverage the benefits of common deep learning accelerators optimized for dense and regular workloads, structured pruning is employed to remove weights in a regular manner (e.g., removing entire channels or layers). 
The LLM Pruner~\cite{ma2023llm} identifies group structures in LLM weights and prunes whole groups, creating a regularly pruned weight matrix, which allows the pruned LLM to run efficiently on GPUs. The Plug-and-Play~\cite{zhang2024plug} prunes weights into a structured N:M sparsity pattern, which can efficiently run on sparse tensor cores in GPUs~\cite{zhu2019sparse}.
SLOPE~\cite{mozaffari2024slope} applies N:M structured sparsity to both forward and backward passes and achieves significant speedups and memory savings compared to dense LLMs.
PGF~\cite{bambhaniya2024progressive} further explored sparse LLM training recipes at high sparsity level, by introducing progressive gradient flow techniques for N:M structured sparsity in transformers, it outperform existing methods on terms of model accuracy.
While structured pruning aligns well with modern GPU requirements, achieving a high compression ratio and maintaining good model performance simultaneously is challenging. 
Unstructured pruning, on the other hand, offers higher flexibility, allowing for better compression ratios and model performance. 
SparseGPT~\cite{frantar2023sparsegpt} achieves up to 50\% weight sparsity in GPT models through optimal partial updates and adaptive masked selection. 
However, while this method reduces memory usage, it may not efficiently reduce computational complexity on common deep learning accelerators optimized for dense workloads. 
Unstructured pruning requires customized hardware support to efficiently utilize sparsity.

To this end, several hardware accelerators have been proposed to efficiently process the sparse matrix multiplication resulting from unstructured weight pruning. 
For instance, the Dual-Side Sparse Tensor Core (DS-STC)~\cite{wang2021dual} modifies tensor core architecture on GPUs to support dual-side sparse matrix multiplication with arbitrary sparsity, outperforming NVIDIA's sparse tensor core on pruned BERT models. 
DS-STC changes the dataflow of the tensor core from inner-product to outer-product, making it more suitable for arbitrary sparse computation. 
Meanwhile, the Row-Merge Sparse Tensor Core (RM-STC)~\cite{huang2023rm} proposes using row merge dataflow for dual-side sparse matrix multiplication, further improving upon DS-STC to achieve high efficiency across all levels of sparsity and reducing the hardware overhead in the design.

\begin{table*}[t] \small
    \caption{LLM accelerators for inference.} 
    \center
    \label{tab:acc}
    \begin{tabular}{cccccc}
      \hline
      Name           &Platform            &Model               &Energy efficiency (TOPS/W) &Quantization/Sparsity & Year\\
      \hline
      TranCIM~\cite{tu202228nm}           &ASIC tapeout 28nm   &BERT          &20.5 (INT8)           &Sparsity    & 2022\\
      DFX~\cite{hong2022dfx}             &FPGA                & GPT-2        &              &            & 2022\\
      X-Former~\cite{xformer}             &PIM simulator 32nm  &BERT          &13.44 (INT8)          & -          & 2022\\
      DOTA~\cite{qu2022dota}              &ASIC 22nm/simulator &GPT-2         & -          &Quantization/Sparsity & 2022\\
      SPRINT~\cite{yazdanbakhsh2022sparse}&PIM simulator 32nm  &GPT-2/BERT    &19.6x                 &Sparsity    & 2022\\
      TransPIM~\cite{zhou2022transpim}    &PIM 65nm/simulator  &GPT-2/BERT    &666.6x RTX 2080Ti     & -          & 2022\\
      Mokey~\cite{mokey}                  &ASIC 65nm/simulator &BERT          &9x GOBO (FP16)        &Quantization& 2022\\
      LeOPArd~\cite{leopard}              &ASIC 65nm/simulator &GPT-2/BERT    &3x SpAtten  &Quantization/Sparsity & 2022\\
      STP~\cite{tambe202322}              &ASIC tapeout 12nm   &BERT          &18.1 (FP4)            &Quantization& 2023\\
      HAIMA~\cite{haima}                  &PIM simulator 45nm  &BERT          & -                    & -          & 2023\\
      TF-MVP~\cite{tf-mvp}                &ASIC 28nm           &BERT/GPT-2    &0.48 (FP16)           &Sparsity    & 2023\\
      TiC-SAT~\cite{tic-sat}              &gem5-X              &BERT          & -                    & -          & 2023\\
      Transformer-OPU~\cite{opu}          &FPGA                &BERT          & -                    & -          & 2023\\
      FACT~\cite{fact}                    &ASIC 28nm           &BERT          &94.88x V100 &Quantization/Sparsity & 2023\\
      TaskFusion~\cite{taskfusion}        &ASIC 22nm/simulator &BERT          &19.83x Jetson Nano    &Sparsity    & 2023\\
      OliVe~\cite{olive}                  &ASIC 22nm/simulator &BERT/GPT-2/OPT&4x GOBO               &Quantization& 2023\\
      C-Transformer~\cite{kim202420}      &ASIC tapeout 28nm   &GPT-2         &33.4 (INT8)           & -          & 2024\\
      SpecPIM~\cite{li2024specpim}        &PIM simulator       &LLaMA/OPT     &6.67x A100 (FP16)     & -          & 2024\\
      ASADI~\cite{asadi}                  &PIM simulator       &BERT/GPT-2    &- (FP32)              &Sparsity    & 2024\\
      AttAcc~\cite{park2024attacc}        &PIM simulator       &LLaMA/GPT-3   &2.67x DGX A100 (FP16) & -          & 2024\\
      NeuPIMs~\cite{heo2024neupims}       &PIM simulator 22nm  &GPT-3         & -                    & -          & 2024\\
      
      \hline
    \end{tabular}
  \end{table*} 
  
\textbf{Sparse Attention.}
Sparse attention is applied to the Multi-Headed Self Attention (MHSA) module in transformers. By limiting the tokens that attend to each other and ignoring the computation of certain attention scores, the complexity and memory access of MHSA are reduced. The sparsity pattern of sparse attention can be defined online or offline, diverging into static sparse attention, which is agnostic to the input data, and dynamic sparse attention, which depends on the input.

Static sparse attention applies pre-defined attention masks to set the corresponding attention scores to zero during inference. The static sparse pattern usually includes local, global, and random attention. 
In local attention, tokens attend only to their neighbors within a fixed window. In global attention, certain tokens attend to all other tokens, regardless of their position. 
In random attention, tokens attend to a set of random tokens, covering various types of dependencies. Longformer~\cite{beltagy2020longformer} utilizes a combination of local attention and global attention to specific tokens, while BigBird~\cite{zaheer2020big} further adds random attention on top of local and global attention, demonstrating its ability to encompass all sequence-to-sequence functions. 
Static sparse attention changes the operations in MHSA from GEMM to SDDMM and SpMM. 
To efficiently perform these operations, sparse NVPIM~\cite{zheng2023accelerating} is proposed to efficiently map sparse attention on processing-in-memory architecture.

For dynamic sparse attention, it removes activations in the attention map according to the value of activations, requiring real-time detection of activations. 
Algorithm and hardware co-design is often used to efficiently determine the sparsity pattern and compute the sparse attention~\cite{taskfusion, asadi, wang2021spatten, qu2022dota, leopard}. 
SpAtten~\cite{wang2021spatten} measures the cumulative importance of tokens or heads and prunes the tokens or heads on the fly. 
The entire token or head is eliminated to preserve a structured sparsity pattern, making computation easier. SpAtten also proposes a parallel top-k engine to identify the sparse pattern. 
DOTA~\cite{qu2022dota} proposes a lightweight detector to omit weak attention score during runtime, inducing a finer-grained sparsity compared with SpAtten and introduces a reconfigurable matrix multiplication unit to cope with the dynamic sparsity pattern.
ASADI~\cite{asadi} introduces a new sparse matrix computation paradigm tailored for the DIA format in self-attention tasks, supported by a highly parallel in-situ computing architecture. 

In summary, sparsification in LLMs, through techniques such as weight pruning and sparse attention, enhances efficiency and reduces computational complexity. However, unlike quantization, the efficiency gains from sparsity are not straightforward and require careful hardware considerations to achieve significant improvements. Unstructured sparsity offers good compression ratios and maintains accuracy, but it necessitates dedicated hardware designs. While proposed solutions for unstructured sparsity are effective, they inevitably introduce additional hardware overhead to manage irregularities. Consequently, in current practices for efficient LLM processing, structured sparsity is often favored. It introduces a degree of regularity that allows for more efficient parallel processing, striking a balance between performance gains and hardware cost.
Sparsity and quantization are usually combined to compress LLMs in practice. \cite{harma2024effective} provides both theoretical and empirical evidence on the optimal way to combine sparsity and quantization in deep neural networks, offering valuable insights for model compression and efficient deployment of large language models.
\subsection{Accelerators for Inference}

The use of LLMs for complex language tasks is exceptionally data- and computation-intensive. 
As a result, there is a strong need for energy-efficient, dedicated processors to minimize these costs, especially on power-constrained edge devices. 
The solutions to achieving this goal and boosting the efficiency of LLM inference involves advancements to both hardware and algorithms, and this research is summarized in Table \ref{tab:acc}. 
Some of these accelerators, which were introduced in the previous subsection, focus on compression techniques such as sparsity and quantization. 
Here, we will present some representative accelerators.

\textbf{Hardware Acceleration.}
On the hardware side, numerous research efforts have been focused on investigating how to take the advantages of novel architectures to minimize costly data movement and enhance computational parallelism. 
For instance, TranCIM~\cite{tu202228nm} follows the non von-Neumann compute-in-memory (CIM) architecture. The digital SRAM-based Bitline-Transpose-CIM macro is introduced to process multiply-accumulate (MAC) operations. By performing MAC operations locally within the SRAM array, CIM macros eliminate excessive and costly data transference for intermediate data. In the TranCIM macros, the SRAM arrays store the weight matrices and take in the input vectors along the bitlines in the same direction. This avoids the need for transpose buffers on the output side to transpose the generated self-attention matrices. Additionally, all the bitlines in each array are activated simultaneously to perform MACs on different weights and inputs, thus improving the computation parallelism. CIM macros that execute different matrix multiplications work in a pipeline manner for further efficiency improvement.

\textbf{Algorithm Acceleration.}
On the algorithm side, optimizing the LLM computation paradigm with compression techniques and the removal of redundant computations enables the LLM processors to achieve both higher efficiency and better hardware utilization. For example, TranCIM supports dynamically selecting dense attention patterns for computation to fully leverage the sparsity in the workload.
Another work, STP~\cite{tambe202322}, exploits the entropy information of input patterns as the criteria to dynamically reconfigure data paths and skip computations of subsequent layers if necessary. This entropy information is also used to customize the local power supply and clock frequency. These techniques lead to a boost in both the throughput and the energy efficiency of the processor with only marginal accuracy loss.
PIVOT~\cite{moitra2024pivot} improves transformer efficiency by dynamically adjusting attention mechanisms based on input complexity, achieving significant reductions in energy consumption.
Finally, C-Transformer~\cite{kim202420} incorporates conventional LLMs with spiking LLMs and executes workloads in both spiking and non-spiking domains to achieve high sparsity as well as high hardware utilization. 

\textbf{Architecture Design.}
Researchers have also proposed accelerators to optimize LLM inference at the architectural level.
Various studies propose architecture designs that facilitate the execution of sparse attention graphs by skipping unnecessary connections. 
Specifically, DOTA~\cite{qu2022dota} introduces a detector for attention selection and utilizes a token-parallel data flow for sparse attention computation, enabling key/value reuse. 
Additionally, SPRINT~\cite{yazdanbakhsh2022sparse} computes attention scores approximately and prunes low attention scores using lightweight analog thresholding circuitry within the processing element (PE) arrays.

Some studies leverage speculative decoding to accelerate LLM inference. 
In speculative decoding, a small draft model generates multiple draft tokens, which are later verified in parallel by the target LLM. 
Meanwhile, SpecPIM~\cite{li2024specpim} finds the optimal resource allocation design through design space exploration, considering the algorithmic and architectural heterogeneity of the draft model and the target LLM.

In addition to digital accelerators, there are efforts to accelerate LLM inference using Processing-In-Memory (PIM) architectures. 
TransPIM~\cite{zhou2022transpim} introduces a token-based dataflow for Transformer-based models, which avoids costly inter-layer data movements. 
Observing that PIM accelerators are more efficient for GEMV computations compared to commercial accelerators like GPUs and TPUs, and that batched decoding alleviates the memory-bound issue of LLM inference on GPUs to some extent, AttAcc~\cite{park2024attacc} and NeuPIMs~\cite{heo2024neupims} propose heterogeneous xPU/PIM systems for batched LLM inference. 
These systems accelerate the attention mechanism on PIM accelerators while assigning other computations to xPUs.

Beyond off-loading scenarios, some studies focus on accelerating LLM inference through distributed systems. 
For instance, DFX~\cite{hong2022dfx} employs model parallelism and an efficient network within a multi-FPGA system, resulting in minimal data synchronization between FPGAs. 
In another study, the authors of CXL-PNM~\cite{park2024lpddr} introduce a processing near memory (PNM) platform using Compute Express Link (CXL), leveraging both model parallelism and data parallelism for workload partitioning.

\subsection{Industry-led AI accelerators}

While numerous hardware and algorithmic accelerations have been proposed in academia, they are limited to certain applications and uses. In the scope of this paper, we are focusing primarily on LLM models such as BERT and GPT-2. This is quite different than industry applications since the target is not only LLMs, but all AI workloads in general. As a result, architecture design must consider more specific requirements. IBM presented RaPiD~\cite{rapid, ibm7nm}, which is a fabricated AI accelerator chip designed for ultra-low precision training and inference. It supports a spectrum of precisions, including the lowest 2-bit fixed point. By utilizing precision scaling, performance and energy improvements are achieved in AI workloads ranging from VGG-16 to BERT. The latest work from IBM, NorthPole~\cite{northpole, ibmscience} differs from their previous works, which were categorized as ASIC accelerators. NorthPole targets large-scale workload, comparing against Google TPUv4~\cite{tpuv4}, NVIDIA A100 \& H100 AI processors. The performance achieved is a joint effort of both architecture and their SDK toolchain. This software-assisted approach is also implemented in their work~\cite{ibm5nm}.

At the same time, Microsoft announced their first in-house AI accelerator, Azure Maia 100, to facilitate their cloud-based AI workloads. It is designed for scalability and sustainability through end-to-end system optimization. It is equipped with a fully custom network protocol and a comprehensive AI framework environment. Cerebras, known for wafer-scale computing, provided a guide for software-hardware co-design for deep learning~\cite{cerebras}. It is clear that LLMs have been developing rapidly from 100 million parameters in BERT to 175 billion parameters in GPT-3 in just a few years. To keep up with the growth of extreme-scale ML models, they proposed a new chip architecture that is wafer-sized.

While numerous LLM accelerators have been proposed in academia, they are mostly for inference and are restricted to certain models. In other words, they are not generalized for different workloads. However, they are able to leverage the unique datapath or observations found in a specific workload to accelerate the computation and achieve power efficiency at the same time. Industry-led AI accelerators, on the other hand, focus on a different perspective. While these accelerations are appreciated, they aren't the general case. Whether it is for edge or cloud-based AI computation, customer targets are very diverse, so accelerators have to be designed to take all AI workloads into consideration. This includes not only inference but also efficient training. Even though the approach for academia and industry is different, both share their critical goal of accelerating LLMs to improve accuracy, power efficiency, latency, and scalability.
\subsection{Other optimizations}
\textbf{Spiking Transformers.}
Merging biologically plausible structures, Spiking Transformers have emerged as an innovative approach to integrating Spiking Neural Networks (SNNs) with Transformer architectures. 
Spiking Transformers have achieved notable advancements in both performance and energy efficiency.
Spikformer \cite{zhou2022spikformerspikingneuralnetwork} was the first to implement spiking self-attention in Transformers. 
It introduced Spikformer, pioneering Spiking Self Attention (SSA) blocks to eliminate resource-intensive multiplications and SoftMax operations. Following this, Masked Spiking Transformers were realized, utilizing Random Spike Masking (RSM) to reduce redundant spikes effectively \cite{Wang2023MaskedSpikingTransformer}. 
Spikingformer demonstrated further innovation \cite{shi2024spikingresformer}, incorporating spike-driven residual learning within Transformer-based SNNs. 
These advancements have been further augmented by the realization of C-Transformer \cite{kim202420}, a processor designed to accelerate Spiking Transformer operations and LLMs. 
It accelerates Spiking Transformers by integrating a Hybrid Multiplication-Accumulation Unit (HMAU), which does accumulation for spiking operations. 
The Output Spike Speculation Unit (OSSU) further enhances efficiency by speculating the output spikes, making the architecture ideal for accelerating spiking neural network operation.

\textbf{Emerging Accelerator Designs.}
The following notable studies contribute unique methodologies to the field of in-memory computing\revref{~\cite{moitra2024trex, ReTransformer, Lu2024RIME, bhattacharjee2024clipformer, insram,linderman2015apparatus, rramnoise, Mohsen2019FloatPIM,  liu2010high, zhao2023raceit}}, specifically showcasing innovations that optimize the efficiency and performance of Transformer models through unique approaches. 
Building on FloatPIM's demonstration of the feasibility of high-precision in-memory acceleration of DNN training~\cite{Mohsen2019FloatPIM}, recent work presented RIME~\cite{Lu2024RIME}, an RRAM-based in-memory floating-point computation architecture aimed at accelerating Transformer inference. 
RIME employs single-cycle NOR, NAND, and innovative minority (Min3) logic functions within RRAM cells to perform high-precision floating-point operations directly in memory. 
Its novel Min3-based adder enables 32-bit floating-point multiplication with minimal cycle count, area and energy consumption.
RACE-IT~\cite{zhao2023raceit} is a reconfigurable analog content-addressable memory and crossbar engine designed for in-memory Transformer acceleration. 
The core innovation is the Compute-ACAM unit, which performs various non-matrix-vector-multiplication operations within the analog domain using analog content-addressable memories (CAMs), significantly improving computation efficiency. 
There is also research about methodologies of in-memory computing for transformers. TReX~\cite{moitra2024trex} proposes a novel approach to optimize Transformers for In-Memory Computing architectures by reusing attention blocks, leading to significant improvements in energy efficiency and area utilization while maintaining high accuracy. ClipFormer~\cite{bhattacharjee2024clipformer} deals with the noise of the memristive crossbar by clipping the KV matrices of Transformers during inference.
These studies highlight recent advancements in in-memory computing architectures for Transformer models.

\section{Conclusion}
\label{sec:conclusion}
This survey has examined the multifaceted challenges and opportunities associated with LLMs. We have delved into various aspects of LLM training, inference, and system integration, highlighting the need for specialized hardware and software solutions. By synthesizing the latest research, we aim to comprehensively understand the trade-offs and design considerations crucial for developing efficient LLM-centric computing systems.
As LLMs continue to evolve and integrate into diverse applications, future research must focus on optimizing their performance and sustainability. This will involve advancing the current methodologies and developing new strategies to enhance their efficiency and practicality. Ultimately, the insights gained from this survey can pave the way for future breakthroughs, enabling the creation of more powerful, efficient, and sustainable LLM systems.

%
\IEEEpeerreviewmaketitle



\ifCLASSOPTIONcaptionsoff
  \newpage
\fi

\bibliographystyle{IEEEtran}
\bibliography{bibAlgo,
bibHW} 



%

%


\begin{IEEEbiography}
[{\includegraphics[width=1in,height=1.25in,clip,keepaspectratio]{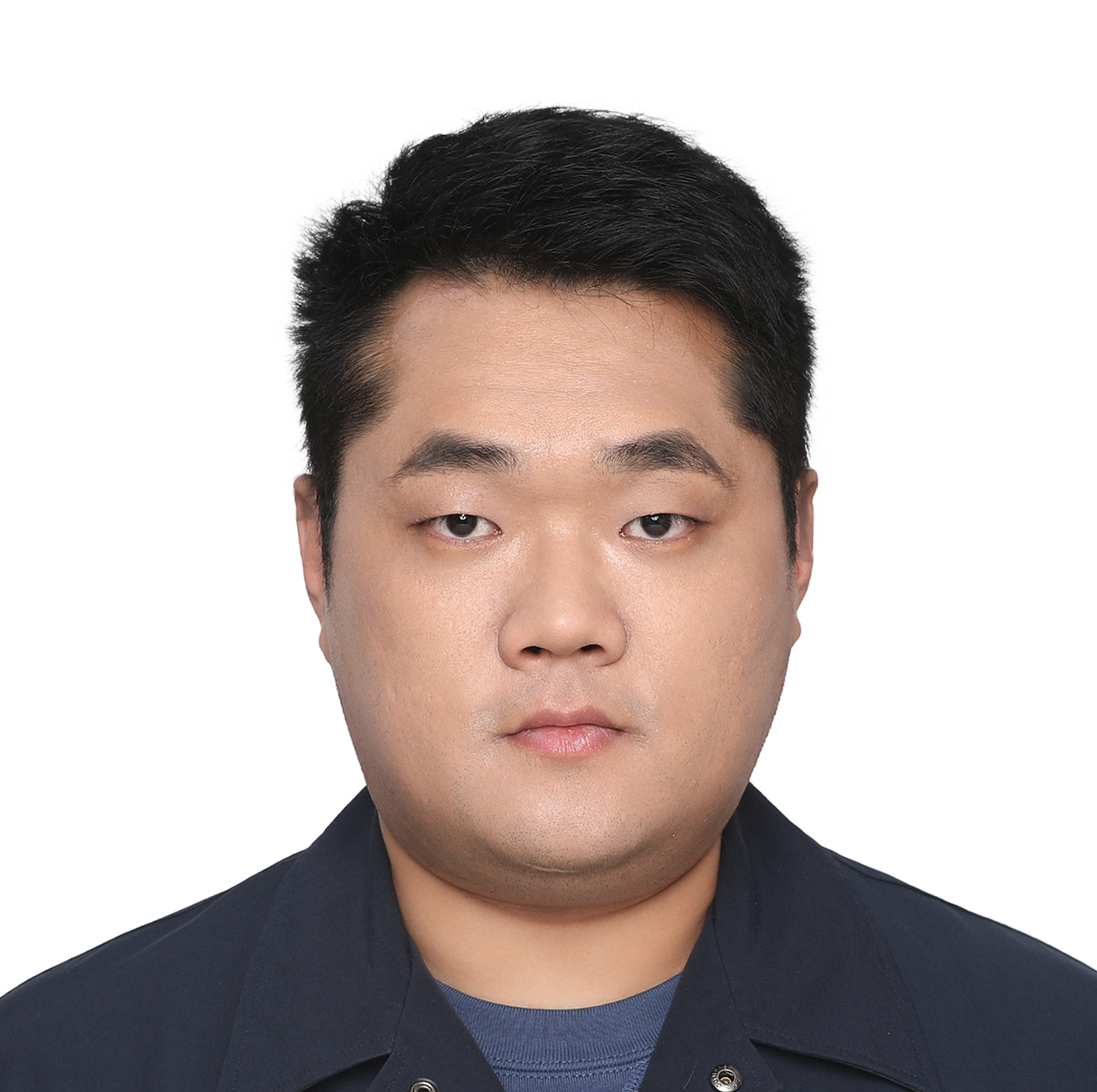}}]
{Cong Guo} received his Ph.D. degree in Computer Science from Shanghai Jiao Tong University, Shanghai, China, in 2023 under the supervision of Prof. Jingwen Leng. 
He is now a Postdoctoral Fellow of Electrical and Computer Engineering at Duke University.
His research interests include computer architecture, high-performance computing, and AI accelerator design.
\end{IEEEbiography}

\begin{IEEEbiography}
[{\includegraphics[width=1in,height=1.25in,clip,keepaspectratio]{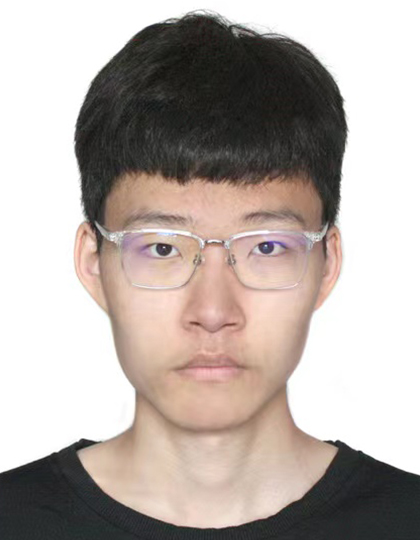}}]{Feng Cheng} received his B.Eng. degree in Electrical Engineering from City University of Hong Kong, Hong Kong, China, in 2022. He is currently pursuing a Ph.D. degree with the Department of Electrical and Computer Engineering, Duke University, Durham, NC, USA, supervised by Prof. Yiran Chen. His research interests include computer architecture, in-/near- memory computing and deep learning accelerator design.
    
\end{IEEEbiography}

\begin{IEEEbiography}
[{\includegraphics[width=1in,height=1.25in,clip,keepaspectratio]{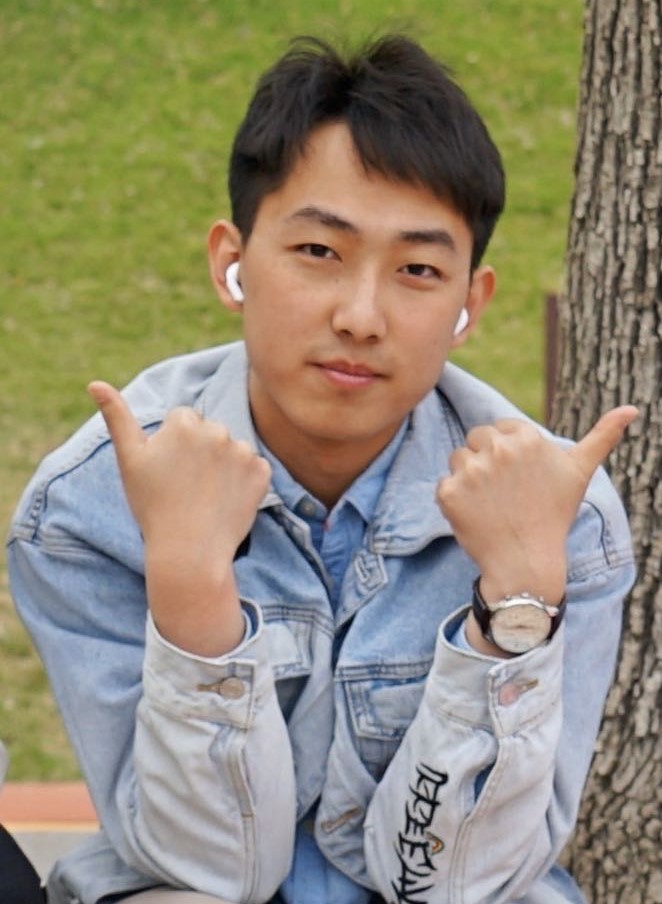}}]{Zhixu Du} received his B.Sc. degree in Mathematics from The University of Hong Kong, Hong Kong, in 2021. He is currently pursuing a Ph.D. degree with the Department of Electrical and Computer Engineering, Duke University, Durham, NC, USA, supervised by Prof. Yiran Chen. His research interests include efficiency, sparsity, fast inference of large models, Mixture-of-Experts, and Federated Learning.
    
\end{IEEEbiography}

\begin{IEEEbiography}
[{\includegraphics[width=1in,height=1.25in,clip,keepaspectratio]{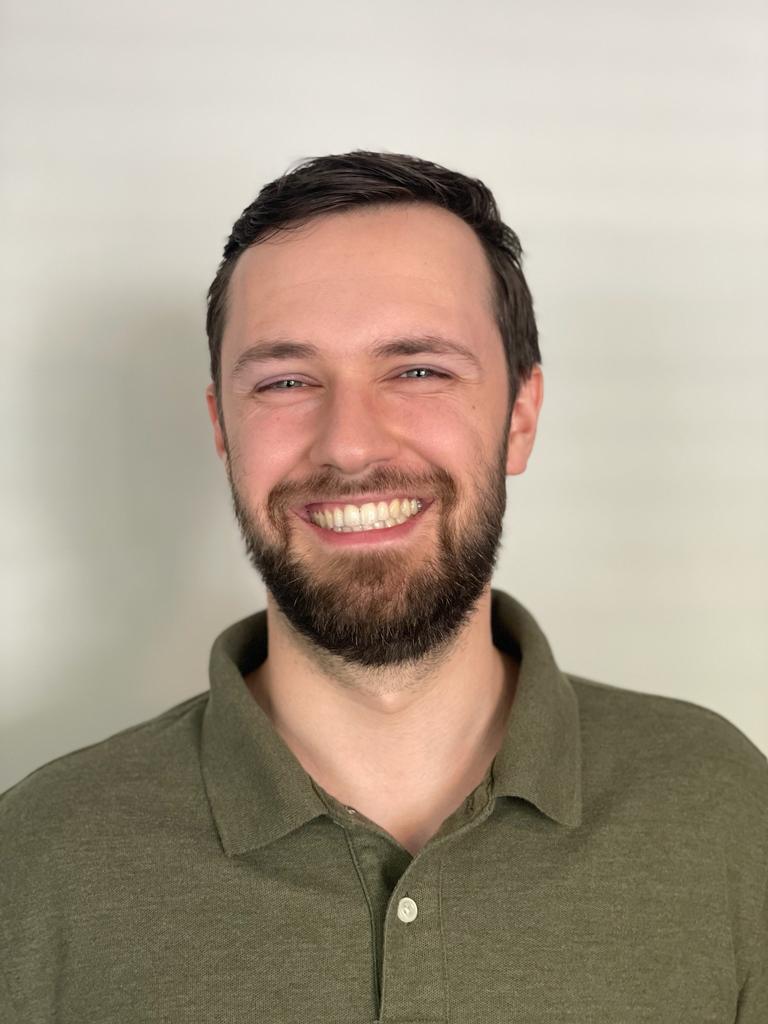}}]
{James Kiessling} received his B.S. degree in Computer Engineering from the University of Rhode Island, Kingston, RI, USA, in 2018. He worked as a firmware engineer until 2023, and is currently pursuing a Ph.D. degree with the Department of Electrical and Computer Engineering, Duke University, Durham, NC, USA, under the supervision of Prof. Yiran Chen. His research interests include efficient deep learning for edge devices, software-hardware co-design for machine learning, and electronic design automation.
\end{IEEEbiography}

\begin{IEEEbiography}[{\includegraphics[width=1in,height=1.25in,clip,keepaspectratio]{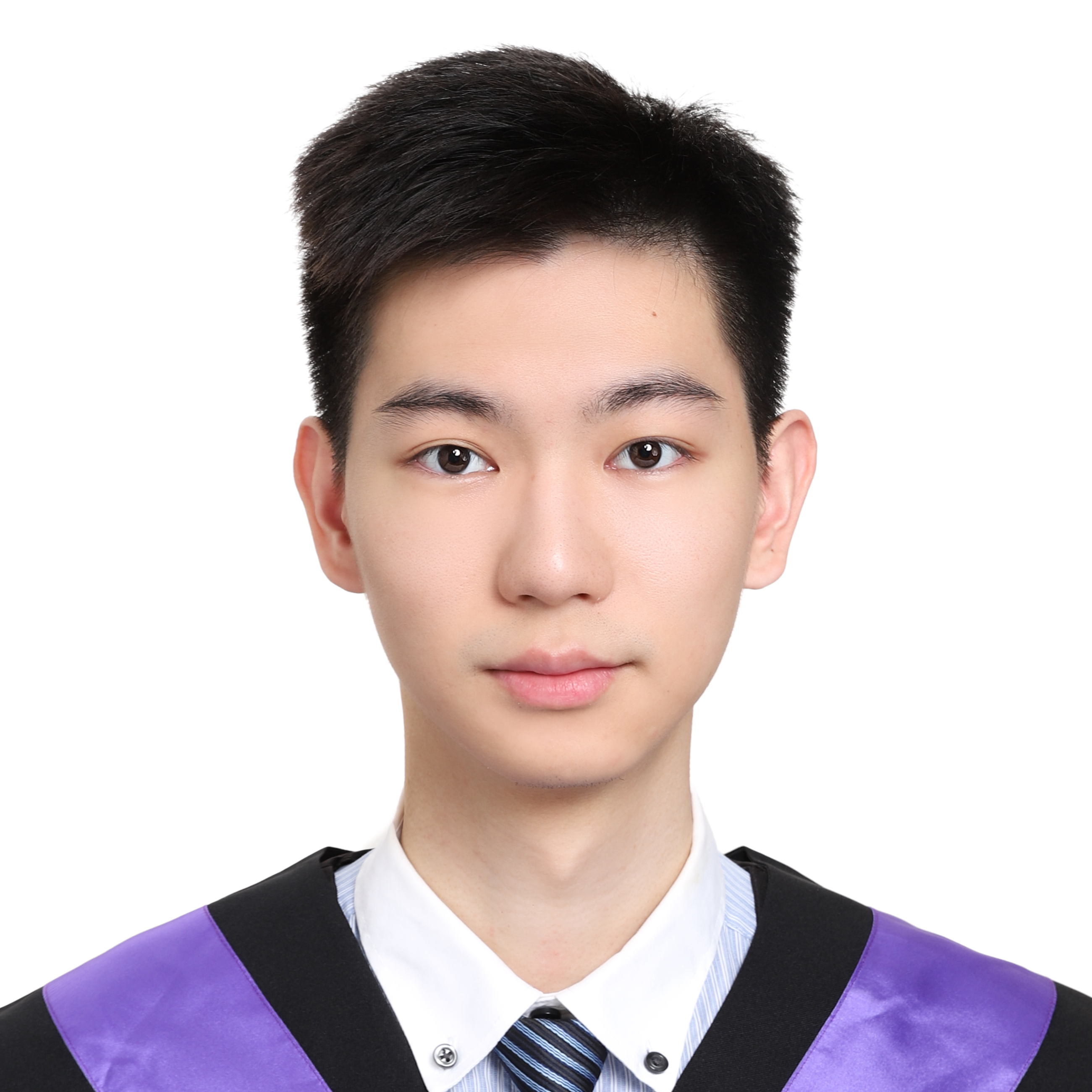}}]{Jonathan Ku} received the B.S. degree in Electrical Engineering and Computer Science from National Tsing Hua University in 2022. He is currently a Ph.D. student in Electrical and Computer Engineering at Duke University, Durham, NC, USA. His research interests include VLSI design, hardware acceleration in cryptography and machine learning.
\end{IEEEbiography}

\begin{IEEEbiography} 
[{\includegraphics[width=1in,height=1.25in,clip,keepaspectratio]{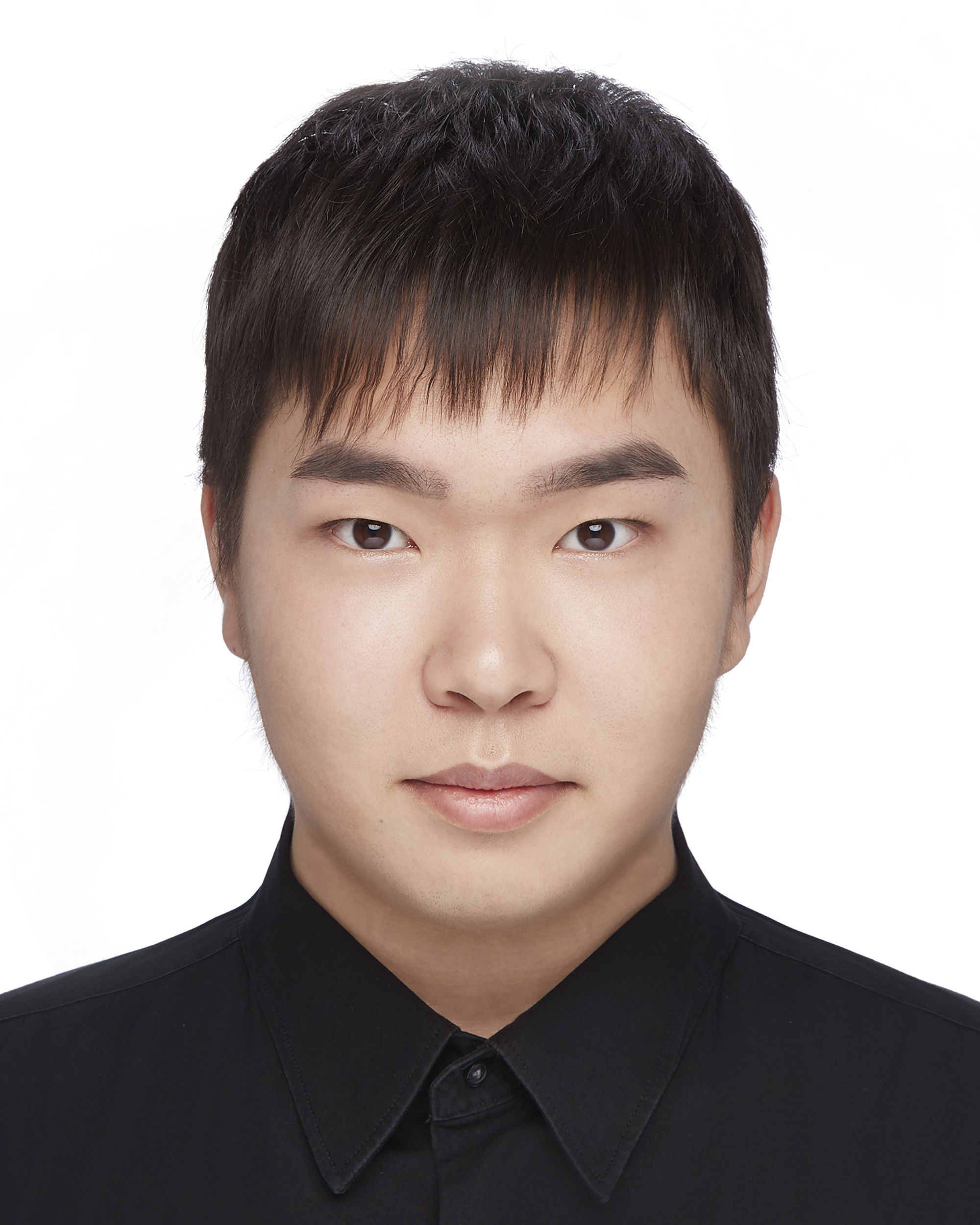}}]{Shiyu Li} received the Ph.D. degree in Computer Engineering from Duke University in 2024 under the guidance of Prof. Yiran Chen. Previously, he received the B.Eng. degree in automation from Tsinghua University, Beijing, China, in 2019. His research interests include computer architecture, algorithm-hardware co-design of deep learning systems, and near-data processing.
\end{IEEEbiography}

\begin{IEEEbiography}[{\includegraphics[width=1in,height=1.25in,clip,keepaspectratio]{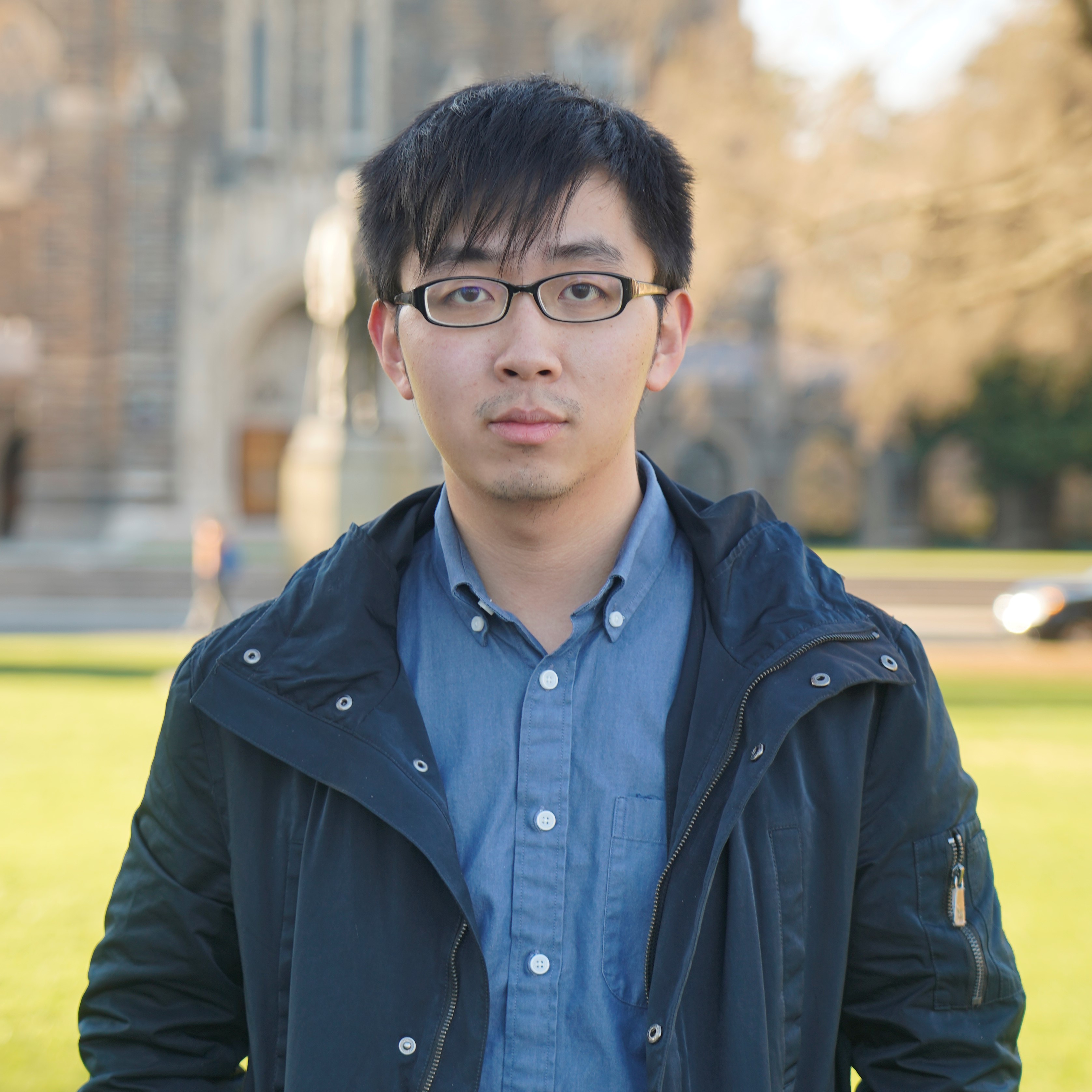}}]{Ziru Li}
received the Bachelor of Engineering (2019) in Electronic Engineering from Tsinghua University, Beijing, China. He received his Ph.D. degree (2024) in Electrical and Computer Engineering at Duke University, Durham, NC, USA, supervised by Prof. Helen Li. His research interests mainly focus on integrated circuit design for advanced artificial intelligence algorithms.
\end{IEEEbiography}

\begin{IEEEbiography}[{\includegraphics[width=1in,height=1.25in,clip,keepaspectratio]{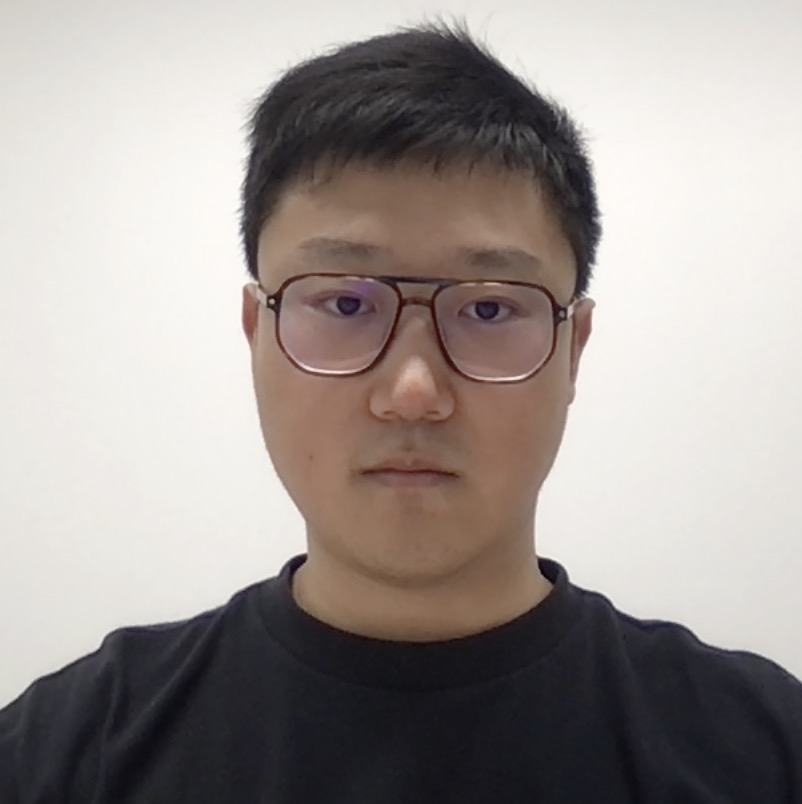}}]{Mingyuan Ma}received the Bachelor of Engineering (2020) in Electronic Engineering from Tsinghua University, Beijing, China. He is currently pursuing his Ph.D. degree in Electrical and Computer Engineering at Duke University, Durham, NC, USA, supervised by Prof. Yiran Chen. His research interests mainly focus on computer architecture and machine learning system.
\end{IEEEbiography}

\begin{IEEEbiography} 
[{\includegraphics[width=1in,height=1.25in,clip,keepaspectratio]{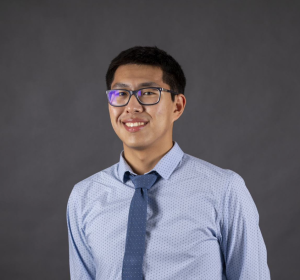}}]{Tergel Molom-Ochir} received his B.S. degree in Electrical Engineering from the University of Massachusetts Amherst, MA, USA, in 2023. He is currently pursuing a Ph.D. degree in the Department of Electrical and Computer Engineering at Duke University, Durham, NC, USA, under the supervision of Prof. Yiran Chen. His research interests include circuit design, in-memory computing, AI accelerators, and non-volatile memories.
\end{IEEEbiography}

\begin{IEEEbiography} 
[{\includegraphics[width=1in,height=1.25in,clip,keepaspectratio]{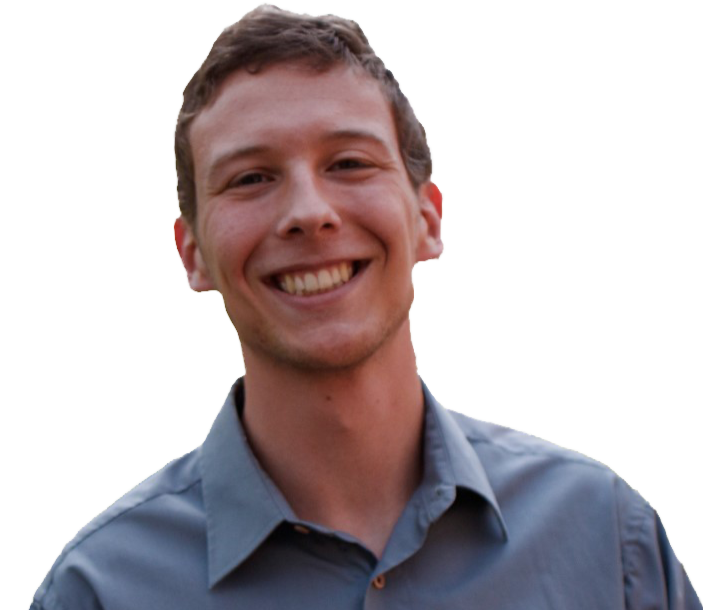}}]{Benjamin Morris} received his B.S. degrees in Computer Science and Biomedical Engineering from North Carolina State University, Raleigh, NC in 2022. He is currently pursuing a Ph.D. degree with the Department of Electrical and Computer Engineering, Duke University, Durham, NC, USA, supervised by Prof. Helen Li. His research interests include computer architecture, algorithm-hardware co-design of data-intensive applications, and near- or in-memory computing.
\end{IEEEbiography}

\begin{IEEEbiography} 
[{\includegraphics[width=1in,height=1.25in,clip,keepaspectratio]{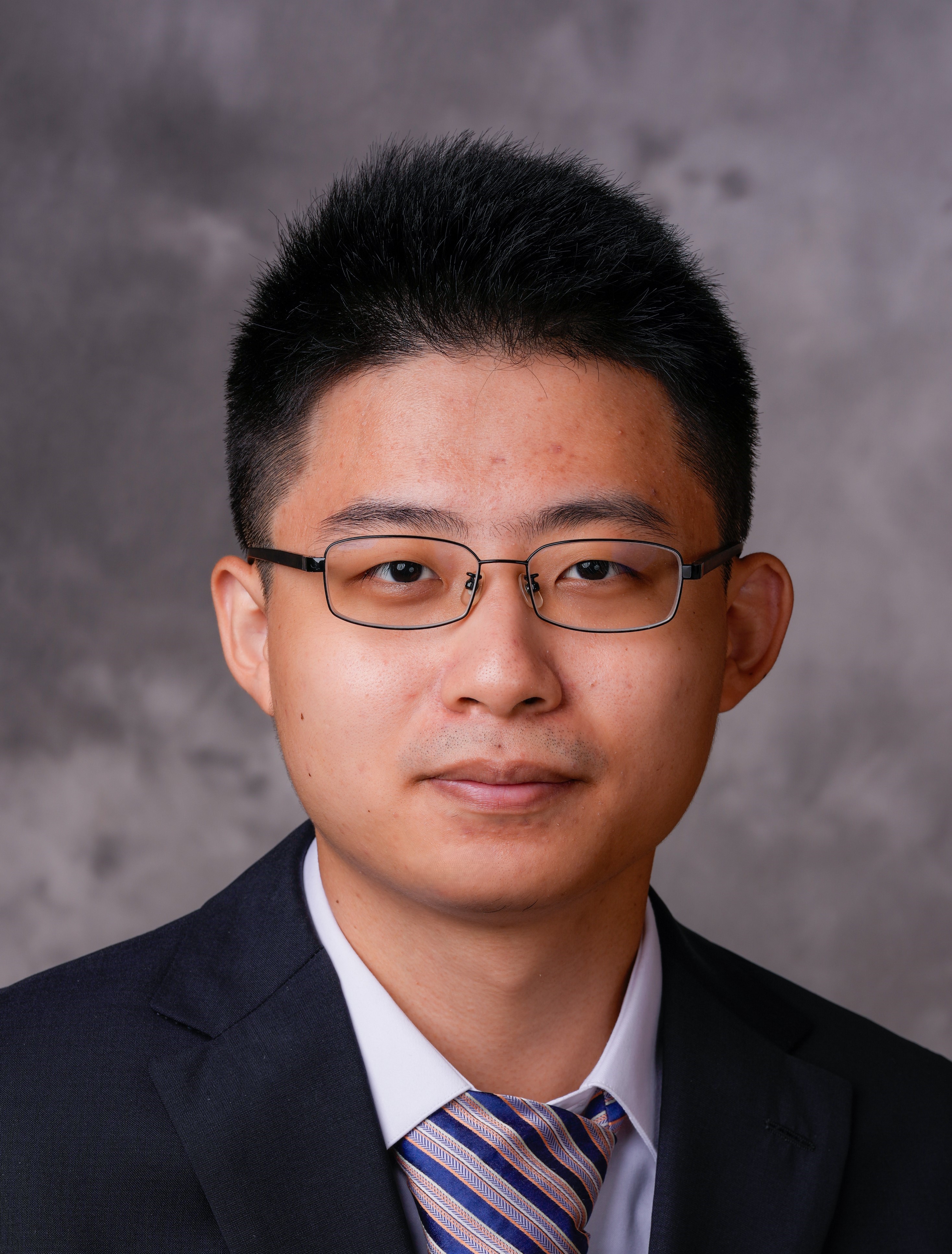}}]{Haoxuan Shan} received his B.S. degree in Electrical and Computer Engineering from Shanghai Jiao Tong University and B.S.E. degree in Computer Science from University of Michigan in 2022. He is currently pursuing a Ph.D. degree with the Department of Electrical and Computer Engineering at Duke University, supervised by Prof. Yiran Chen. His research interests include computer architecture and algorithm-hardware co-design. 
\end{IEEEbiography}

\begin{IEEEbiography} 
[{\includegraphics[width=1in,height=1.25in,clip,keepaspectratio]{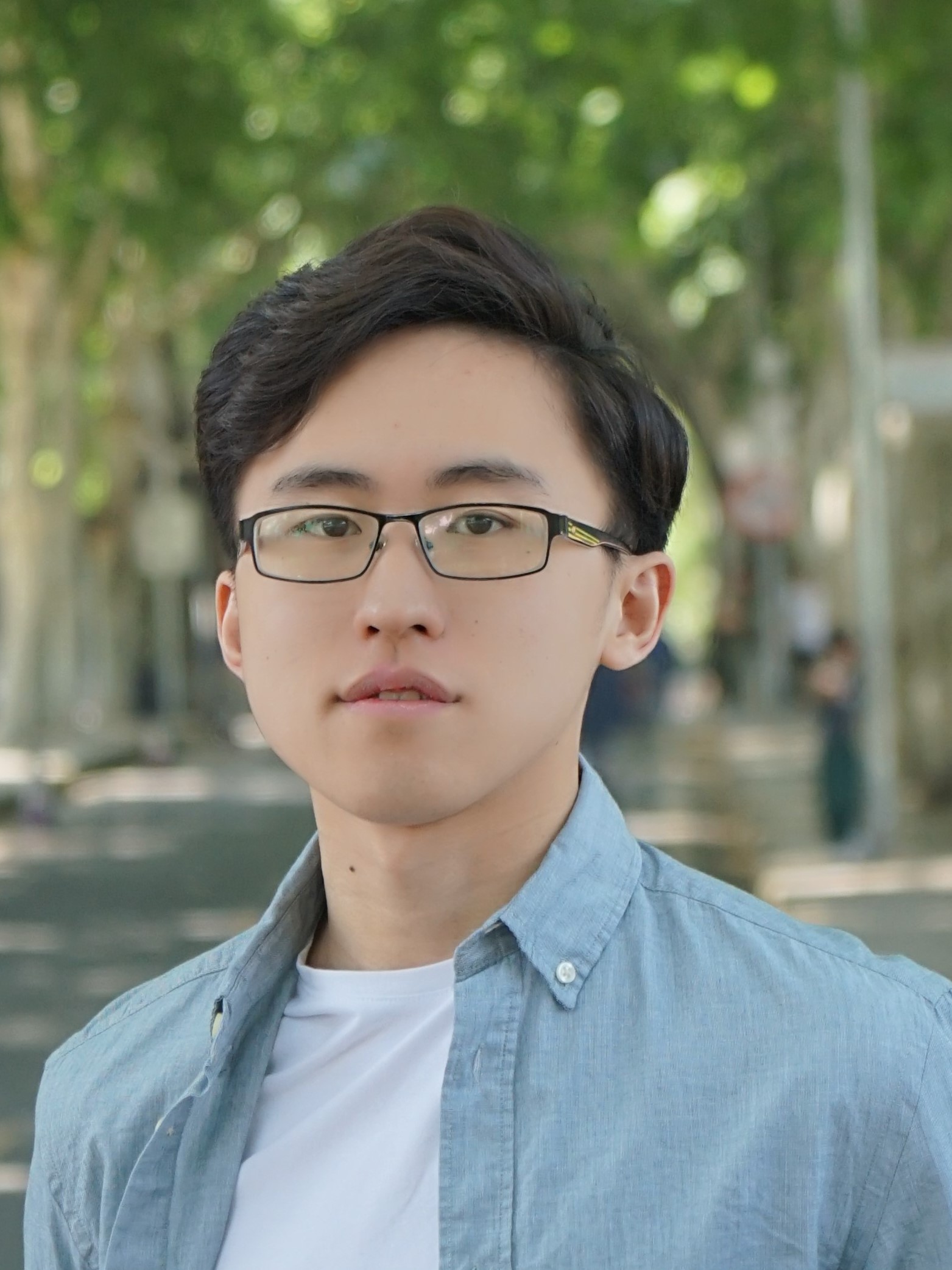}}]{Jingwei Sun} received his B.E. degree in Electrical Engineering from Wuhan University in 2019 and M.S. degree in Electrical and Computer Engineering from Duke University in 2021. He is currently pursuing a Ph.D. degree with the Department of Electrical and Computer Engineering at Duke University, supervised by Prof. Yiran Chen. His research interests include machine learning systems and edge computing. 
\end{IEEEbiography}

\begin{IEEEbiography}
[{\includegraphics[width=1in,height=1.25in,clip,keepaspectratio]{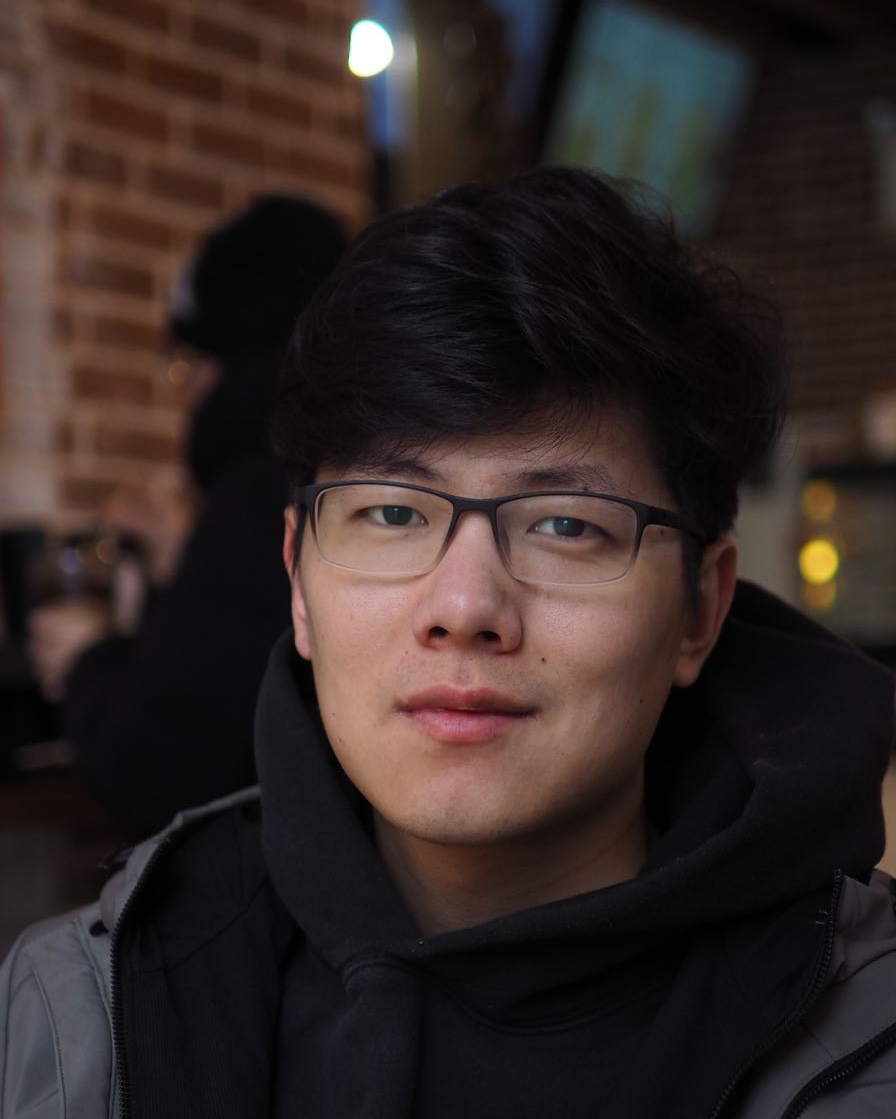}}]{Yitu Wang} received the B.Eng. degree in Micoelectronics from Fudan University, Shanghai, China, in 2020. He is currently pursuing a Ph.D. degree with the Department of Electrical and Computer Engineering, Duke University, Durham, NC, USA, supervised by Prof. Yiran Chen. His research interests focus on the near storage/memory architecture design for data-intensive applications and deep learning system.
    
\end{IEEEbiography}

\begin{IEEEbiography}
[{\includegraphics[width=1in,height=1.25in,clip,keepaspectratio]{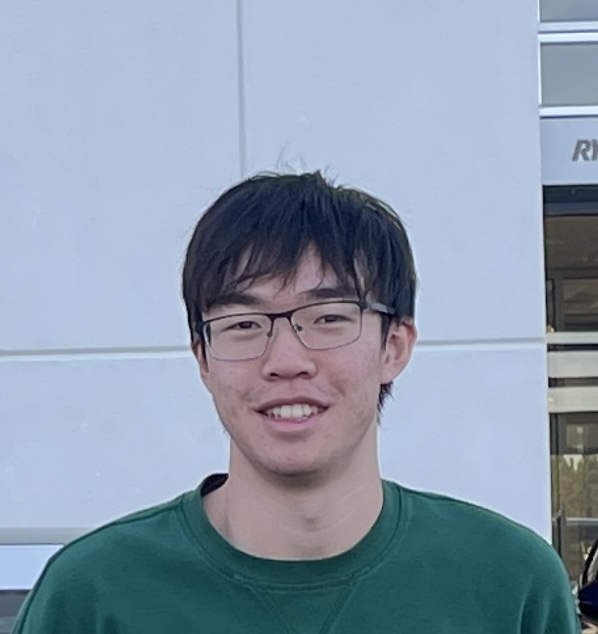}}]{Chiyue Wei} received his B.Eng. degree in Electronic Engineering from Tsinghua University, Beijing, China, in 2023. He is currently pursuing a Ph.D. degree with the Department of Electrical and Computer Engineering, Duke University, Durham, NC, USA, supervised by Prof. Yiran Chen. His research interest include computer architecture and software-hardware co-design for machine learning.
\end{IEEEbiography}

\vspace{-5mm}

\begin{IEEEbiography}[{\includegraphics[width=1in,height=1.25in,clip,keepaspectratio]{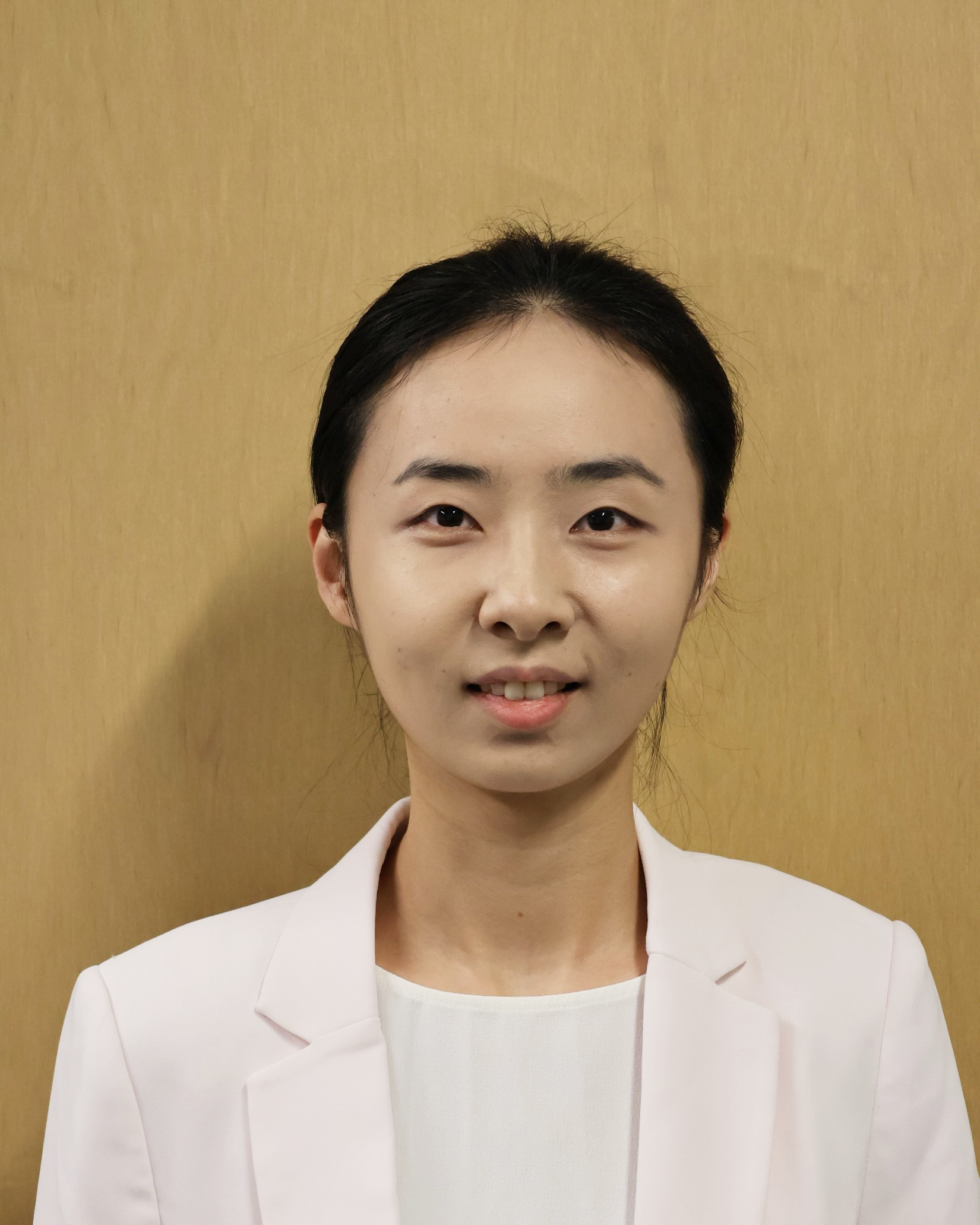}}]{Xueying Wu} received her B.Eng. degree in Microelectronics from Fudan University, Shanghai, China, in 2021. She is currently pursuing a Ph.D. degree with the Department of Electrical and Computer Engineering, Duke University, Durham, NC, USA, supervised by Prof. Hai (Helen) Li. Her research interests include computer architecture, neural network compression, and hardware-software co-design of Processing-in-Memory systems.
\end{IEEEbiography}

\vspace{-5mm}

\begin{IEEEbiography}
[{\includegraphics[width=1in,height=1.1in,clip,keepaspectratio]{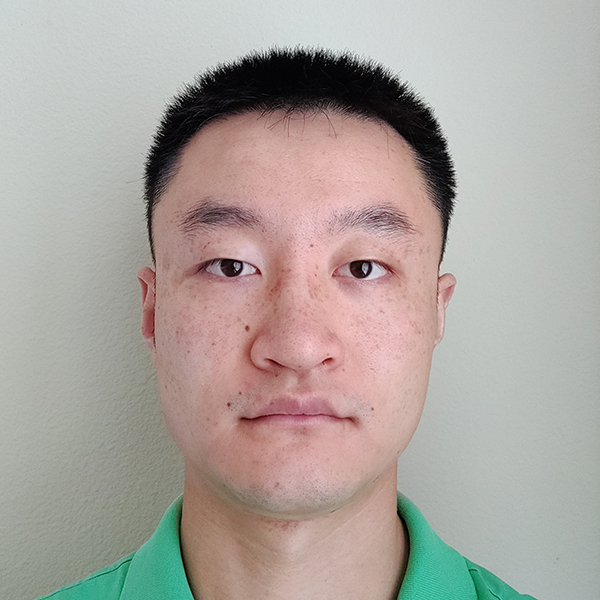}}]{Yuhao Wu} received his B.S. and M.S. in Computer Science from the University of Arkansas, Fayetteville, AR, USA, in 2015. He worked as a software engineer until 2021, and is currently pursuing a Ph.D. degree in the Department of Electrical and Computer Engineering, Duke University, Durham, NC, USA, under the supervision of Prof. Yiran Chen. His research interests include Federated Learning and medical image/video analysis.
\end{IEEEbiography}

\vspace{-5mm}
\begin{IEEEbiography}[{\includegraphics[width=1in,height=1.1in,clip,keepaspectratio]{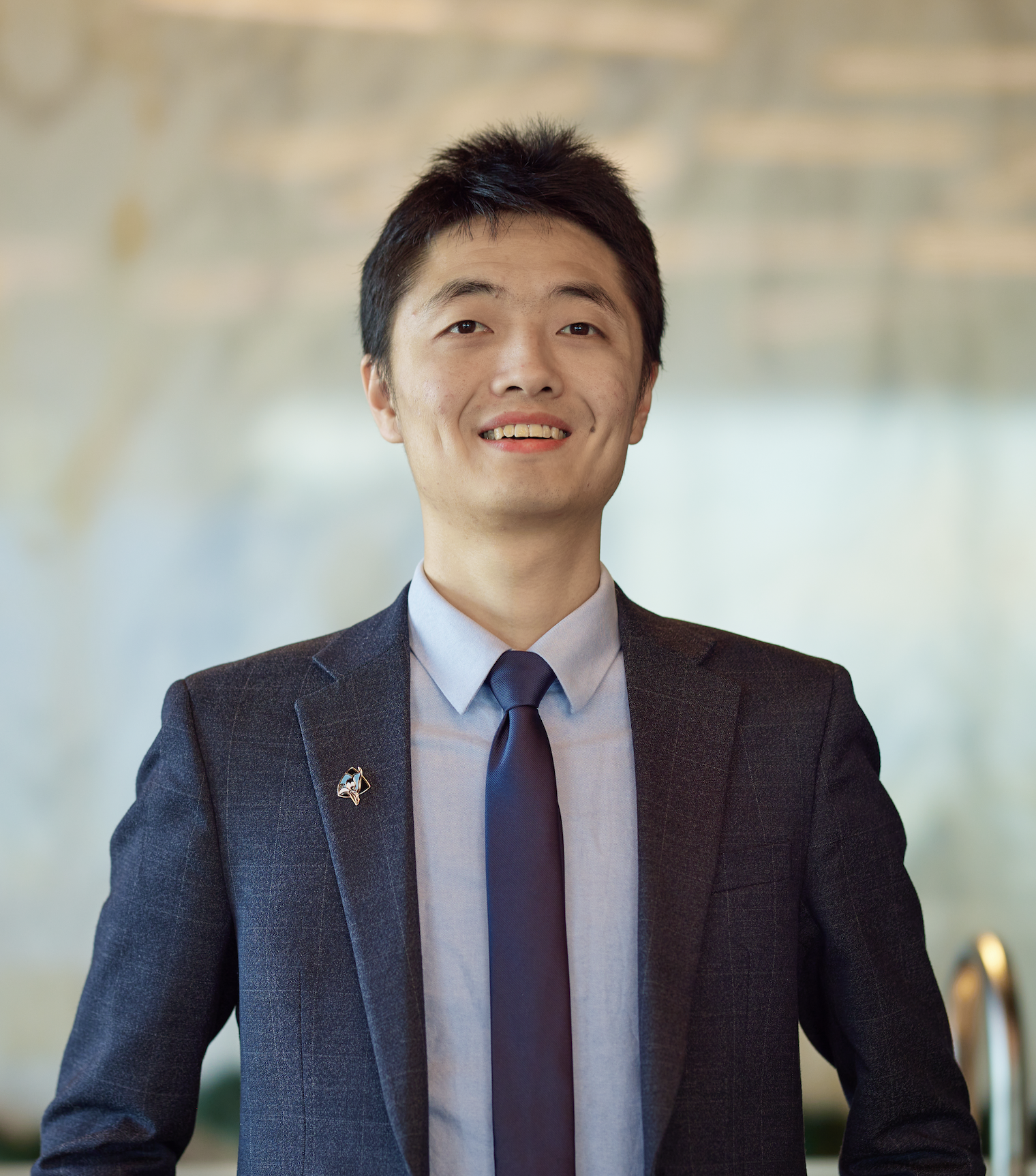}}]{Dr. Hao (Frank) Yang}
 received B.S. degrees in Telecommunication Engineering from Beijing University of Posts and Telecommunications and the University of London in 2017, the Ph.D. degree in Civil Engineering from the University of Washington, and finished his post-doc at Duke University. He is currently an assistant professor at Johns Hopkins University, Department of Civil and System Engineering. His research focuses on developing trustworthy machine learning and data science methods to improve the equity, safety, resilience, and sustainability of traffic systems. Dr. Yang has published over 20 top refereed journal papers and 45 conference proceedings, with three best paper awards. 
\end{IEEEbiography}

\vspace{-5mm}
\begin{IEEEbiography}
[{\includegraphics[width=1in,height=1.25in,clip,keepaspectratio]{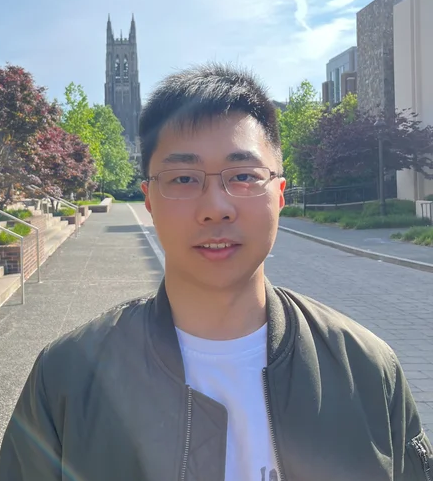}}]
{Jingyang Zhang} received his B.Eng. degree in Electronic Engineering from Tsinghua University, Beijing, China, in 2019.
Since then, he has been pursuing the Ph.D. degree at Dept. of Electrical and Computer Engineering at Duke University, supervised by Dr. Yiran Chen and Dr. Hai (Helen) Li.
His research spans from robustness of deep learning-based vision systems to (more recently) generative AI and multi-modal LLMs.
\end{IEEEbiography}

\begin{IEEEbiography}
[{\includegraphics[width=1in,height=1.25in,clip,keepaspectratio]{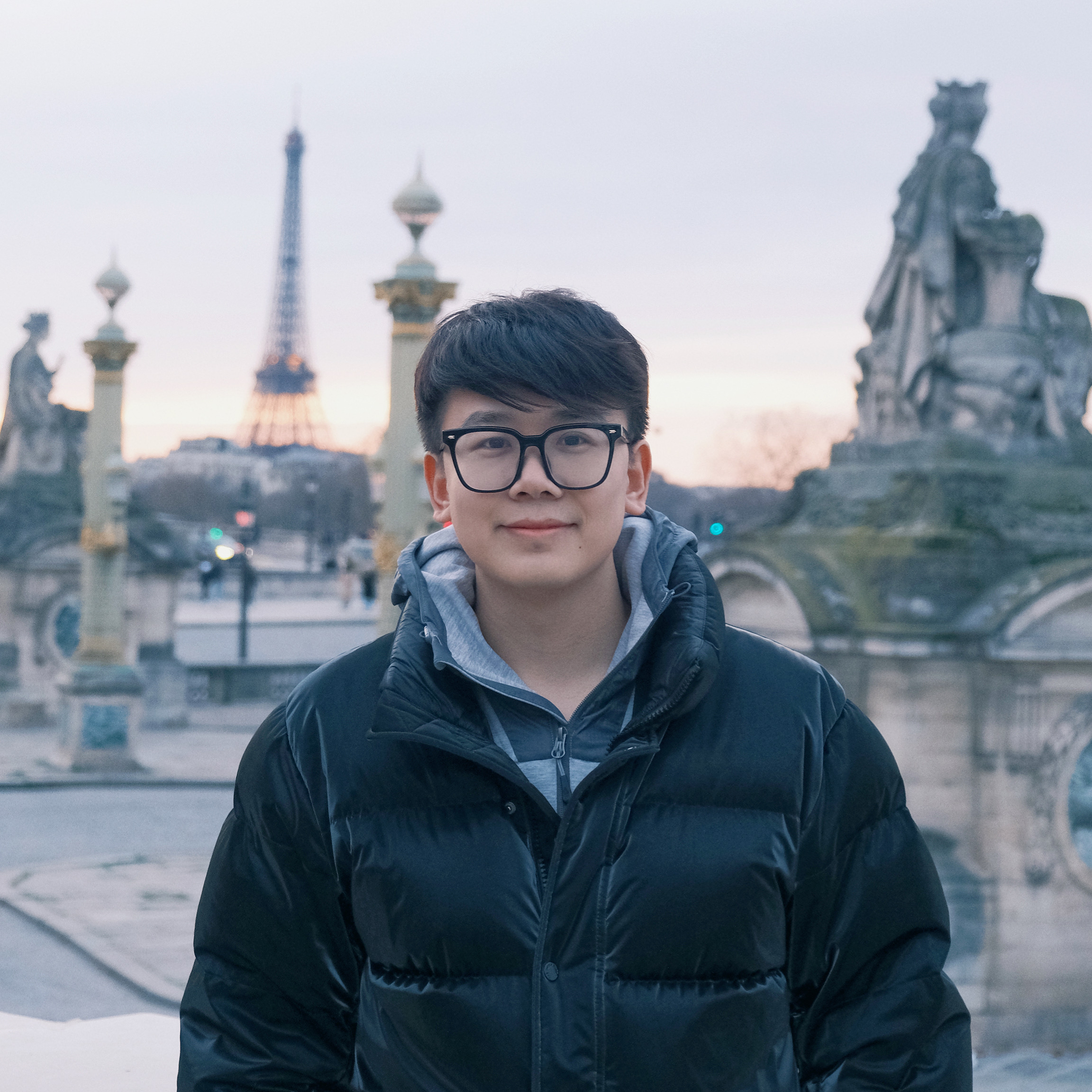}}]{Junyao Zhang} received his M.S. in Electrical Engineering from University of Southern California, Los Angeles, USA in 2021. He is currently pursuing a Ph.D. degree in the Department of Electrical and Computer Engineering at Duke University, Durham, USA, under the supervision of Prof. Yiran Chen. His research interests include electronic design automation, quantum computing and high-performance computing.
\end{IEEEbiography}

\vspace{-5mm}
\begin{IEEEbiography}
[{\includegraphics[width=1in,height=1.25in,clip,keepaspectratio]{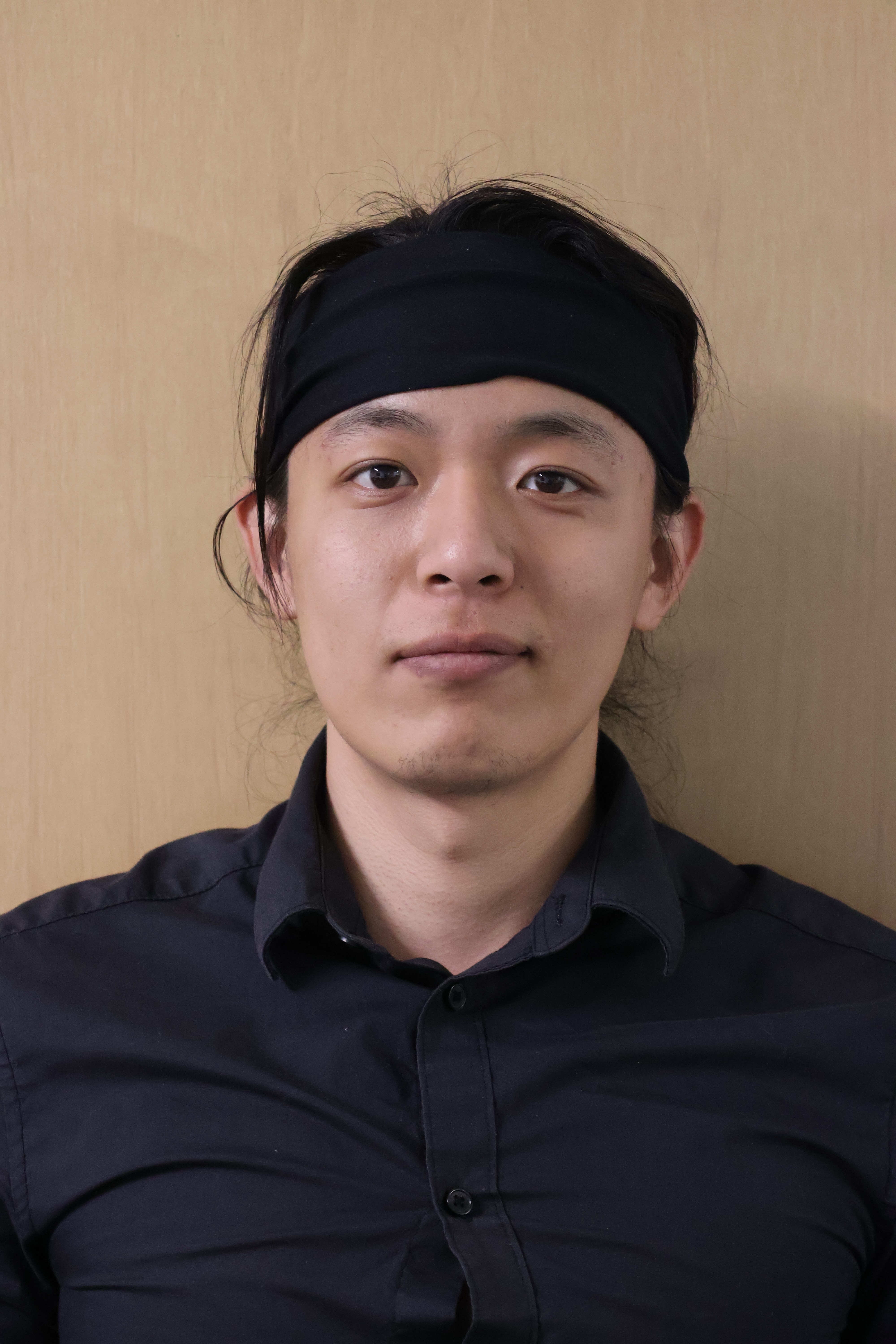}}]{Qilin Zheng}  received his Ph.D degree (2024) in Electrical and Computer Engineering at Duke University. 
Before that, he received his M.Sc and B.Sc degree from KU Leuven and Peking University, respectively. His research interests include machine learning accelerator, in-memory computing and non-volatile memory design. 
\end{IEEEbiography}

\vspace{-5mm}
\begin{IEEEbiography}
[{\includegraphics[width=1in,height=1.25in,clip,keepaspectratio]{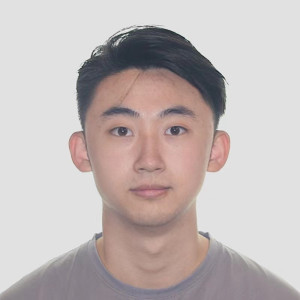}}]{Guanglei Zhou} received his B.Eng. degree in Electrical Engineering from City University of Hong Kong, Hong Kong, China, in 2020 and a M.A.Sc degree in Computer Engineering from University of Toronto, Toronto, Canada in 2022. He is currently pursuing a Ph.D. degree with the Department of Electrical and Computer Engineering, Duke University, Durham, NC, USA, supervised by Prof. Yiran Chen. His research interests include computer architecture and electronic design automation.
\end{IEEEbiography}

\vspace{-5mm}
\begin{IEEEbiography}[{\includegraphics[width=1in,height=1.25in,clip,keepaspectratio]{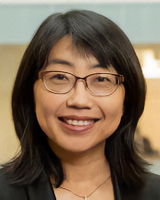}}]{Hai (Helen) Li (M'08-SM'16-F'19)}
received the Ph.D. degree from Purdue University in 2004. 
Dr. Li is currently the Clare Boothe Luce Professor and Department Chair of the Electrical and Computer Engineering Department at Duke University.
Her current research interests include neuromorphic circuits and systems for brain-inspired computing, machine learning acceleration and trustworthy AI, conventional and emerging memory design and architecture, and software and hardware co-design.
Dr. Li is a recipient of the NSF Career Award (2012), DARPA Young Faculty Award (2013), TUM-IAS Hans Fischer Fellowship from Germany (2017), and ELATE Fellowship (2020). She received 9 best paper awards and additional 9 best paper nominations from international conferences. Dr. Li is a Distinguished Lecturer of the IEEE CAS Society (2018-2019) and a Distinguished Speaker of ACM (2017-2020). 
\end{IEEEbiography}

\vspace{-5mm}
\begin{IEEEbiography}
    [{\includegraphics[width=1in,height=1.25in,clip,keepaspectratio]{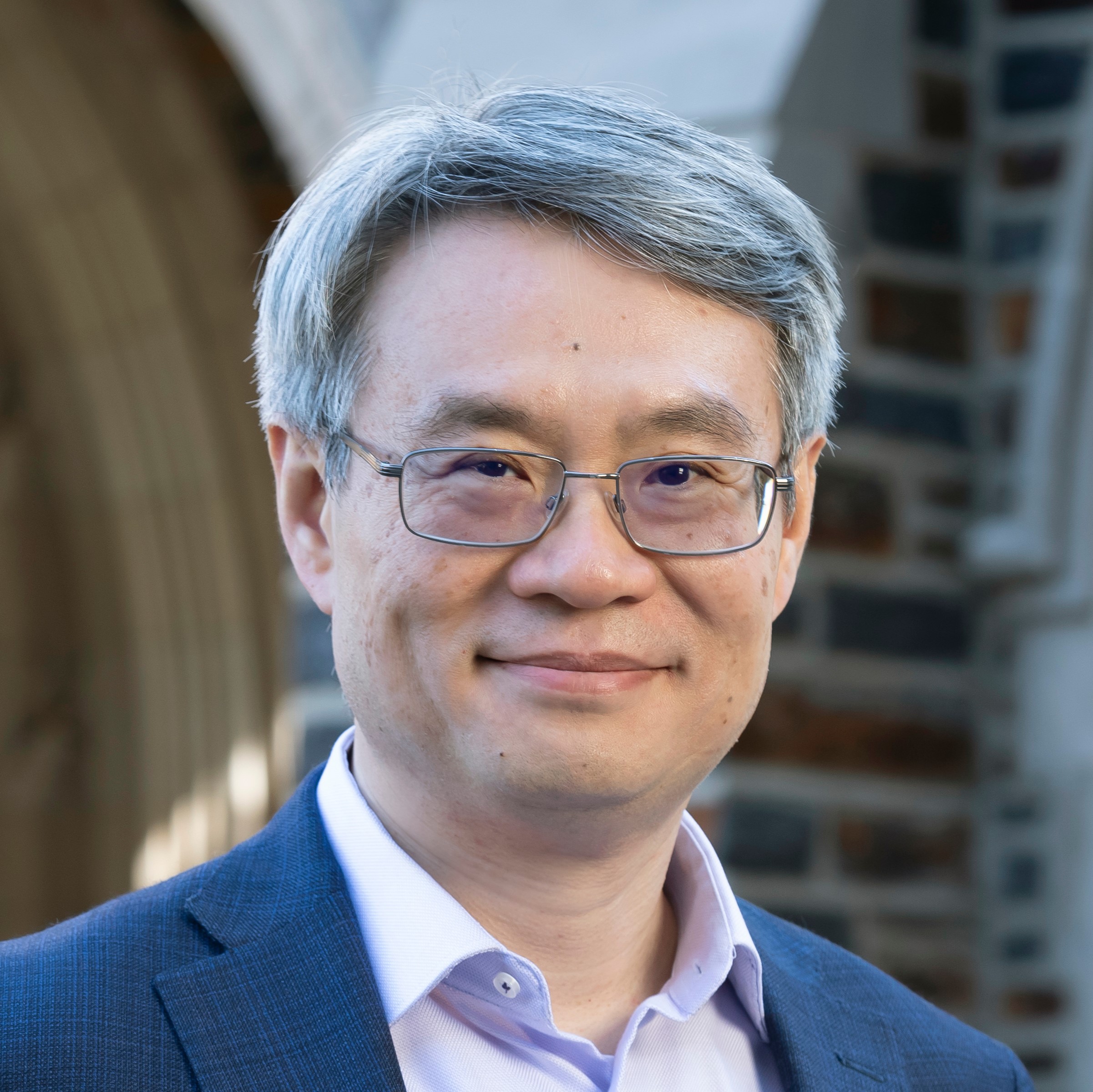}}]{Yiran Chen}  received the Ph.D. degree from Purdue University in 2005. 
He is now the John Cocke Distinguished Professor of Electrical and Computer Engineering at Duke University and serving as the director of the NSF AI Institute for Edge Computing Leveraging the Next-generation Networks (Athena), the NSF Industry-University Cooperative Research Center (IUCRC) for Alternative Sustainable and Intelligent Computing (ASIC), and the co-director of Duke Center for Computational Evolutionary Intelligence (DCEI).
He received 11 best paper awards, 1 best poster award, and 15 best paper nominations from international conferences and workshops. 
He received numerous awards for his technical contributions and professional services such as the IEEE CASS Charles A. Desoer Technical Achievement Award, the IEEE Computer Society Edward J. McCluskey Technical Achievement Award, etc. He has been the distinguished lecturer of IEEE CEDA and CAS.
\end{IEEEbiography}




\end{document}